\documentclass[prd,onecolumn,superscriptaddress,groupedaddress,nofootinbib]{revtex4}  

\usepackage{amssymb,amsmath,amsfonts,amsbsy}
\usepackage{color}
\usepackage{enumerate}
\usepackage{graphicx} 
\usepackage{hyperref}

\hypersetup{
    colorlinks=true,
    linkcolor=red,
    citecolor=blue,
} 

\newcommand{\al}{\alpha} 
\newcommand{\kp}{\kappa}
\newcommand{\psip}{\psi'} 
\newcommand{\apsip}{A_{\psi'}} 
\newcommand{\bpsip}{B_{\psi'}}

\newcommand{\be}{\begin{equation}} 
\newcommand{\ee}{\end{equation}} 
\newcommand{\bea}{\begin{eqnarray}}
\newcommand{\eea}{\end{eqnarray}}

\begin{document}

\title{The Well-Tempered Cosmological Constant}

\author{Stephen Appleby}\email{stephen@kias.re.kr}
\affiliation{School of Physics, Korea Institute for Advanced Study, 85
Hoegiro, Dongdaemun-gu, Seoul 02455, Korea}
\author{Eric V.~Linder}\email{evlinder@lbl.gov} 
\affiliation{Berkeley Center for Cosmological Physics \& Berkeley Lab, 
University of California, Berkeley, CA 94720, USA}  
\affiliation{Energetic Cosmos Laboratory, Nazarbayev University, 
Astana, Kazakhstan 010000}

\begin{abstract} 
Self tuning is one of the few methods for dynamically cancelling a large cosmological constant and yet 
giving an accelerating universe. Its drawback is that it tends to screen all sources of 
energy density, including matter. We develop a model that tempers the self tuning so the 
dynamical scalar field still cancels an arbitrary cosmological constant, including the vacuum 
energy through any high energy phase transitions, without affecting the matter fields. 
The scalar-tensor gravitational action is simple, related to cubic Horndeski gravity, with 
a nonlinear derivative interaction plus a tadpole term. Applying shift symmetry and using the 
property of degeneracy of the field equations we find families of functions that admit de Sitter 
solutions with expansion rates that are independent of the magnitude of the cosmological constant 
and preserve radiation and matter dominated phases. That is, the method can deliver a standard cosmic history including current acceleration, despite the presence of a Planck scale cosmological constant. 
\end{abstract}

\date{\today} 

\maketitle

\section{Introduction} 

Vacuum energy exists. Even apart from current cosmic acceleration (see e.g.\ \cite{Copeland:2006wr,Joyce:2014kja,Joyce:2016vqv,Huterer:2017buf} for some reviews), particle physics predicts that vacuum 
(zero-point) energy should be associated with fields in the universe. At the same time, general relativity allows for a 
cosmological constant within the field equations. The observation of cosmic acceleration indicates that there 
is something like a vacuum energy (with equation of state such that its effective pressure $P$ satisfies $P \simeq -\rho$ where $\rho$ is its effective energy density), at a level far below the usual high energy physics expectations. 
Thus we seem to need two occurrences: removal or cancellation of the high energy physics scale cosmological 
constant (or at least its gravitational effects), and appearance of a vacuum energy at a scale (energy density 
$\sim$ meV$^4$) corresponding to current cosmic acceleration. 

The first property -- ``the original cosmological constant problem'' -- in particular is challenging, especially 
when taking into account fields entering at different energies as the universe evolves (i.e.\ one cannot just 
cancel the vacuum energy once), and quantum radiative corrections (a vacuum energy cancelled at tree level may 
not stay cancelled when quantum interactions are included). A substantial literature on this exists -- see 
\cite{Weinberg:1988cp,Martin:2012bt} for a review of the issues and \cite{Nobbenhuis:2004wn,Padilla:2015aaa,ArkaniHamed:2002fu,Kaloper:2013zca,Kaloper:2014dqa,Kaloper:2014fca, Charmousis:2011bf,Charmousis:2011ea} for some novel attempts at resolving them -- and two of the most promising ideas seem to be self tuning for dynamically cancelling vacuum 
energy and shift symmetry for controlling quantum corrections. Moreover, one must not only cancel the cosmological 
constant from the early universe and replace it with a low energy vacuum energy for current acceleration, but 
ensure a viable cosmic history with radiation and matter in between. 

Self tuning from certain actions giving nonlinear field equations of dynamical scalar fields \cite{Charmousis:2011bf,Charmousis:2011ea} works well at the first two requirements, but tends to fail at the last -- it tunes away {\it all\/} forms of 
energy density \cite{Linder:2013zoa}. We present here a new self tuning solution that only screens vacuum 
energy, leaving radiation and matter untouched. Two key ingredients are utilizing the property of degeneracy in 
the field equations, and using shift symmetry in the field. This delivers the harmonious tuning fulfilling all the 
requirements in the early, the radiation/matter dominated, and the current cosmic acceleration epochs, so we call 
it a ``well tempered'' theory. 

In Section~\ref{sec:review} we review the concept of self tuning with a 
basic example. Section~\ref{sec:method} introduces the 
degenerate equations approach and we put forward an action with all the desired properties. We work through several examples and show their approach to 
a de Sitter state, with cancelling a high energy cosmological constant, 
analytically. Furthermore we impose constraints from freedom from ghosts 
and instability in Section~\ref{sec:ghost}. 
Section~\ref{sec:numerical} presents numerical solutions, 
verifying the analytic results. We demonstrate the preservation of matter 
dominated intermediate epochs, transitioning to late time acceleration, in 
Section~\ref{sec:matter}. We conclude in Section~\ref{sec:concl} and give 
some other examples of action functions for self-tuning in Appendix~\ref{sec:apx}, and compare to another degeneracy approach in Appendix~\ref{sec:appsfe}.

\section{\label{sec:review}Review of Self Tuning Scalar Fields}

We begin by briefly reviewing the concept of self tuning. The progenitor of this class of scalar-tensor self tuning models is the Fab Four \cite{Charmousis:2011bf,Charmousis:2011ea}. However, in \cite{Charmousis:2011bf,Charmousis:2011ea} Minkowski space solutions were studied. As we are primarily interested in the existence of de Sitter states, in this section we review a different (but closely related) model \cite{Gubitosi:2011sg,Appleby:2011aa,Appleby:2012rx} that will serve as a closer analogy to the action considered in the rest of the article. 

We study the following scalar tensor action, 

\begin{equation}\label{eq:acex} S = \int d^{4}x \sqrt{-g}\, \left[ {M_{\rm pl}^{2} \over 2}R + c_{2}X + {c_{\rm G} \over M^{2}} G^{\mu\nu}\nabla_{\mu}\phi \nabla_{\nu}\phi \right] + S_{\rm m}[g_{\mu\nu},\Psi_{i}] \ , \end{equation} 
where $R$ is the Ricci scalar, $M$ is a mass scale, $c_{2}, c_{\rm G}$ are order unity dimensionless constants, $G^{\mu\nu}$ is the Einstein tensor, $\phi$ is an additional scalar field derivatively coupled to the metric, $X = -\nabla_{\alpha}\phi \nabla^{\alpha}\phi/2$, $M_{\rm pl}$ is the Planck mass and $S_{\rm m}$ is the action of matter fields $\Psi_{i}$. We assume that $\phi$ is not explicitly coupled to matter. 

We calculate the field equations taking a flat, FLRW metric 

\begin{equation}\label{eq:flrw} ds^{2} = -dt^{2} + a^{2}(t) \delta_{ij} dx^{i}dx^{j} \ ,\end{equation} 
and time dependent scalar field $\phi(t)$, arriving at the following field equations describing the background expansion,  

\begin{eqnarray}\label{eq:fried} & & 3 M_{\rm pl}^{2} H^{2} = \rho_{\Lambda} + \rho_{\rm m} + {c_{2} \over 2} \dot{\phi}^{2} + 9 {c_{\rm G} \over M^{2}} H^{2} \dot{\phi}^{2} = \rho_{\Lambda} + \rho_{\rm m} + \rho_{\phi} \\ 
\label{eq:hdotcg} & & 2 \left( M_{\rm pl}^{2} - {c_{\rm G} \over M^{2}} \dot{\phi}^{2} \right) \dot{H} = \rho_{\Lambda} -P_{\rm m} - 3 M_{\rm pl}^{2}H^{2} -{c_{2} \over 2} \dot{\phi}^{2} + 3 {c_{\rm G} \over M^{2}}H^{2} \dot{\phi}^{2} + 4 {c_{\rm G} \over M^{2}} H \dot{\phi}\ddot{\phi}  \\ 
\label{eq:sfe} & & \left( c_{2} + 6 {c_{\rm G} \over M^{2}} H^{2} \right) \ddot{\phi} = -3H \left( c_{2} + 6 {c_{\rm G} \over M^{2}} H^{2} \right) \dot{\phi} - 12 {c_{\rm G} \over M^{2}} \dot{\phi} H \dot{H} \\ 
& & \dot{\rho}_{\rm m} + 3H \left( \rho_{\rm m} + P_{\rm m} \right) = 0 \ , \end{eqnarray} 
where we have included a generic matter component with density $\rho_{\rm m}$ and pressure $P_{\rm m}$, and a constant vacuum energy $\rho_{\Lambda}$. For this particular model there exists a `pseudo' fixed point with constant Hubble expansion rate, 

\begin{equation}\label{eq:h0} H_{\rm ds}^{2} = -\frac{c_{2}}{6c_{\rm G}}M^{2} \ , \end{equation} 
which is independent of $\rho_{\Lambda}$. One of the crucial ingredients of self-tuning is that for this de Sitter state the scalar field equation ($\ref{eq:sfe}$) is trivially satisfied. It follows that the scalar field $\phi$ does not relax to a constant value, but rather evolves with dynamics described by 
Equations~($\ref{eq:fried},\,\ref{eq:hdotcg}$). The condition that the scalar field equation is trivially satisfied is only true {\it on-shell}, that is for $H = H_{\rm ds}$. 

Weinberg's no-go theorem \cite{Weinberg:1988cp} informs us that for a generic classical Lagrangian density ${\cal L}(g_{\mu\nu},\Psi_{i})$ with matter fields $\Psi_{i}$, any vacuum state in which all fields relax to constant vacuum expectation values will require fine tuning to eliminate the vacuum energy. A key component of this no-go theorem is that the vacuum is translationally invariant, and hence all fields must be constant on-shell. 
This condition is violated by the theory described by the action ($\ref{eq:acex}$), giving the loophole for self tuning, for the vacuum state ($\ref{eq:h0}$) as the scalar field $\phi$ evolves at this de Sitter point. 

The key ingredient that allows the scalar field to dynamically cancel the vacuum energy in this model is {\it degeneracy} of the field equations -- we are demanding a solution of the form $H={\rm const} = H_{\rm ds}$ exists regardless of the vacuum and scalar field energy densities $\rho_{\Lambda}$, $\rho_{\phi}$. Applying the ansatz $H=H_{\rm ds}$, we can solve the Hamiltonian constraint ($\ref{eq:fried}$) for $\phi$. The system (\ref{eq:fried}--\ref{eq:sfe}) is then overconstrained, unless the scalar field equation is redundant at this point. This is achieved in the standard self-tuning mechanism by enforcing that the scalar field equation ($\ref{eq:sfe}$) is trivially satisfied at the de Sitter point, and generates a time dependent $\phi(t)$ solution at $H=H_{\rm ds}$. In summary, a necessary (but not sufficient) requirement for self-tuning to occur is that the field equations must exhibit redundancy at the vacuum state \cite{Charmousis:2011bf,Charmousis:2011ea}. 

In spite of its many interesting properties, numerous works have highlighted issues with the model that cast doubt on its viability to describe the dynamics of the universe \cite{Starobinsky:2016kua, Niedermann:2017cel}. The main issues are:  

\begin{enumerate} 
\item{The scalar field not only tunes the effect of the vacuum energy $\rho_{\Lambda}$ but also any other matter and radiation that is present \cite{Linder:2013zoa}. Specially chosen scalar field potentials for self-tuning models can generate epochs of expansion that mimic standard matter and radiation eras \cite{Copeland:2012qf, Martin-Moruno:2015lha}. Stability of the background spacetime is also an issue \cite{Starobinsky:2016kua}.}

\item{Application of linear perturbation theory indicates that the graviton propagator will be significantly modified by the presence of the scalar field and local fifth force constraints are likely to be violated \cite{Niedermann:2017cel}. This is a generic problem with light scalar fields \cite{Chiba:2006jp}, and can be evaded by a non-linear screening mechanism \cite{Vainshtein:1972sx, Khoury:2003rn, Hinterbichler:2010es}. For a model to be considered viable it must proved that a non-linear solution exists that forces the graviton propagator close to its general relativistic form in certain regimes (small scales or high density).}

\item{Recent gravitational wave experiments   \cite{TheLIGOScientific:2017qsa} with electromagnetic follow-up \cite{,Coulter:2017wya,GBM:2017lvd,Murguia-Berthier:2017kkn} have revealed that the graviton effectively propagates at the speed of light \cite{Monitor:2017mdv} which poses a serious challenge to scalar-tensor theories \cite{Lombriser:2015sxa,Lombriser:2016yzn,Bettoni:2016mij, Creminelli:2017sry,Sakstein:2017xjx,Ezquiaga:2017ekz, Baker:2017hug,Arai:2017hxj,Battye:2018ssx}. Kinetic mixing of $\phi$ and $g_{\mu\nu}$ of the form $G^{\mu\nu}\nabla_{\mu}\phi \nabla_{\nu}\phi$ is known to modify the speed of the graviton, leading to apparent inconsistency with observations  . This narrative has been challenged in \cite{Babichev:2017lmw}, where it was argued that a disformal transformation can be applied to fix the graviton speed to $c_{\rm grav} = c_{\rm light}$ and a class of self-tuning models that generalise the action ($\ref{eq:acex}$) remains viable.}

\end{enumerate}

In this work, we wish to study an alternative class of models that might evade these three issues. In what follows, we will take the positive properties of such self-tuning models and create a 
well tempered theory that can potentially ameliorate each of the negative aspects.

\section{From Self Tuning to Well Tempering} \label{sec:method} 

To satisfy the gravitational wave speed constraint we give up on the 
$G^{\mu\nu}\nabla_{\mu}\phi \nabla_{\nu}\phi$ form but use a different non-linear term that can 
also potentially satisfy the fifth force screening constraint.  In brief, we study the scalar-tensor action 
\begin{equation}\label{eq:scal1} S = \int d^{4}x \sqrt{-g} \left[ {M_{\rm pl}^{2} \over 2} R + K(\phi,X) - G(\phi,X)\Box \phi + \rho_{\Lambda} \right] \ , \end{equation} 
where $\rho_\Lambda$ is the bare, high energy physics cosmological constant 
energy density. This form 
has been considered within the context of dark energy \cite{Deffayet:2010qz} and inflation \cite{Kobayashi:2011nu}. 
Here $K, G$ are arbitrary functions of $\phi$ and $X$. We will later impose  shift symmetry of the scalar field. The action~($\ref{eq:scal1}$) is a member 
of the Horndeski class of scalar-tensor theories   \cite{Horndeski:1974wa,Nicolis:2008in,Deffayet:2009wt,Deffayet:2011gz}, where we have neglected all other terms to preserve the gravitational wave speed equal to the speed of light. Initially we 
fix the matter contribution to be zero for simplicity, including only a vacuum energy 
$\rho_{\Lambda}$; we re-introduce matter in Section~\ref{sec:matter}.

\subsection{Field Equations and Degeneracy} 

For the flat FLRW metric ($\ref{eq:flrw}$) we can write the Hamiltonian constraint, scale factor, and scalar field dynamical equations as \cite{Kobayashi:2010cm, Kobayashi:2011nu} 
\begin{eqnarray} \label{eq:s6_1} & & 3 M_{\rm pl}^{2} H^{2} = \rho_{\Lambda} + 2K_{X}X - K + 3G_{X} H \dot{\phi}^{3} - 2 G_{\phi}X \\ 
\label{eq:s6_2} & & -M_{\rm pl}^{2} \left( 3H^{2} + 2 \dot{H} \right) = p_{\Lambda} + K - 2 \left( G_{\phi} + G_{X} \ddot{\phi}\right) X \\
 & & \nonumber K_{X} \left( \ddot{\phi} + 3 H \dot{\phi}\right) + 2 K_{XX}X\ddot{\phi} + 2 K_{X\phi}X - K_{\phi} - 2 \left( G_{\phi} - G_{X\phi}X\right)\left( \ddot{\phi} + 3H\dot{\phi}\right) + \\ 
\label{eq:s6_3} & & \qquad \qquad + 6 G_{X}\left[ (HX)\dot{} + 3 H^{2} X\right] - 4 G_{X\phi}X\ddot{\phi} - 2 G_{\phi\phi} X + 6 H G_{XX}X \dot{X} = 0 \end{eqnarray}

\noindent To preserve the shift symmetry $\phi \to \phi + c$ for constant $c$, we take the following functional forms: 
\bea  
K(\phi, X) &=& c_{1}M^{3} \phi + M^{4} A(X) \ , \label{eq:kx}\\ 
G(\phi,X) &=& M B(X) \ , \label{eq:gx} 
\eea 
for arbitrary dimensionless functions $A(X)$, $B(X)$, and where 
$c_1$ is a constant. Note the 
tadpole term proportional to $\phi$ is shift symmetric as the 
constant offset can be absorbed into the cosmological constant. 

Rewriting the field equations in terms of dimensionless parameters  
\begin{equation} \psi = {\phi \over M},  \qquad h = {H \over M}, \qquad \tau = Mt\ ,  \end{equation} 
with $\rho_{\Lambda} = M_{\rm pl}^{2} M_{\Lambda}^{2}$ and 
$X=M^4 (\psi')^{2}/2$, we arrive at 
\begin{eqnarray} 
 3{M_{\rm pl}^{2} \over M^{2}} h^2&=& {M_{\rm pl}^{2} M_{\Lambda}^{2} \over M^{4}} +\psi'\apsip-c_1\psi-A+3h\psip^2\bpsip \label{eq:h2}\\ 
2{M_{\rm pl}^{2} \over M^{2}}h'\,f(\psi')&=&-\apsip\psi'f(\psi') +\bpsip\psi'\,(\psi''-3h\psi')f(\psi') \label{eq:hdot}\\ 
0&=&3h\apsip+A_{\psi' \psi'}\psi''-c_1+3\bpsip\,(h'\psi'+h\psi''+3h^2\psi')+3hB_{\psi' \psi'}\psi'\psi'' \ , 
\label{eq:psi2} 
\end{eqnarray} 

\noindent where we have used the Hamiltonian constraint ($\ref{eq:s6_1}$) in Equation~(\ref{eq:s6_2}). Primes denote derivatives with respect to dimensionless time $\tau$ and $\psi'$ subscripts indicate derivatives with respect to $\psi'$. We have multiplied the scale factor equation~(\ref{eq:hdot}) by an arbitrary, non-zero function of $\psi'$, $f(\psi')$. The field equations (\ref{eq:h2}--\ref{eq:psi2}) give the general description of the dynamics of this system\,\footnote{In 
terms of the property functions of \cite{Bellini:2014fua}, we have 
\begin{eqnarray} 
\al_T&=&0\ , \\ 
\al_M&=&0\ , \qquad M_\star^2=M_{\rm pl}^2\ , \\ 
\al_B&=&(2\dot\phi/H)XG_X=(M/M_{\rm pl})^2(\psi')^3\bpsip/h \ , \\ 
\al_K&=&\frac{2X}{H^2}(K_X+2XK_{XX})+12\frac{\dot\phi}{H}X(G_X+XG_{XX})=\frac{M^2}{h M_{\rm pl}^2}(\psi')^2A_{\psi'\psi'} +6\frac{M^2}{h M_{\rm pl}^2}(\psi')^3B_{\psi'\psi'} \ . 
\end{eqnarray} 
Note that the important tadpole term $c_1$ does not enter into the property functions, only appearing 
explicitly in the background equations, suitable for tuning away a cosmological constant.}. 

We now wish to move {\it on-shell\/} and search for a particular 
solution to the equations for which there is a de Sitter state 
$H=H_{\rm ds}$, that is $h=\kp \equiv H_{\rm ds}/M = {\rm const}$. We stress that the 
following equations apply only {\it on-shell}, that is after imposing the ansatz $h=\kappa$. For this choice, 
Equations~(\ref{eq:hdot}, \ref{eq:psi2}) reduce to 
\begin{eqnarray} 
0&=&-\apsip F(\psi') +\bpsip\,(\psi''-3\kp\psi')F(\psi') \label{eq:hdotds}\\ 
0&=&3\kp\apsip+A_{\psi' \psi'}\psi''-c_1+3\bpsip\,(\kp\psi''+3\kp^2\psi')+3\kp B_{\psi' \psi'}\psi'\psi'' 
\label{eq:psi2ds} \ , 
\end{eqnarray} 
where $F(\psi') \equiv \psi' f(\psi') \ne 0$. 

Demanding the existence of a solution $h=\kappa$ that is independent of $\psi$ and $\rho_{\Lambda}$ generically overconstrains the dynamical system. The standard self tuning mechanism \cite{Charmousis:2011bf,Charmousis:2011ea} evades this issue by requiring that the scalar field  
equation~($\ref{eq:psi2ds}$) 
is identically zero {\it on-shell\/}, and also possesses a non-trivial $h'$ dependence away from the de Sitter state. In Appendix~\ref{sec:appsfe} we study the class of functions $A,\,B$ that generate self-tuning solutions using this standard mechanism. However, there is an alternative form of degeneracy that we focus on in this work. Specifically, we search for functions $A,B$ for which Equations~(\ref{eq:hdotds}, \ref{eq:psi2ds}) 
are equivalent. 

To enforce this condition we separately equate the coefficients of the $\psi''$ terms, and all other terms. That is, we demand that the functions $A$, $B$ satisfy the following conditions 
\begin{eqnarray}\label{eq:con1} & & F(\psi') B_{\psi'} =   A_{\psi'\psi'} + 3 \kp \left( B_{\psi'\psi'}\psi' + B_{\psi'}\right)  \\  \label{eq:con2} & & - F(\psi') \left( A_{\psi'} + 3 \kp B_{\psi'}\psi' \right) = 3 \kp \left( A_{\psi'} + 3 \kp B_{\psi'}\psi' \right) - c_{1} \ .\end{eqnarray}
We can partially solve these equations and express $A_{\psi'}$, $B_{\psi'}$ in terms of $F$ as 
\begin{eqnarray}\label{eq:at1} & & A_{\psi'} = {c_{1} \over 3 \kp +F} + 3c_{1} \kp\, {F_{\psi'}\psi' \over F(3\kp +F)^{2}} \\
\label{eq:bt1} & & B_{\psi'} = - c_{1}\,  {F_{\psi'} \over F(3\kp +F)^{2}} \ . \label{eq:bpsip} 
\end{eqnarray} 
Note that both $A$ and $B$ require $c_{1} \ne 0$, i.e.\ the tadpole term plays a crucial role. 

Plugging this partial solution back into the {\it on-shell\/} Equations~(\ref{eq:hdotds}, \ref{eq:psi2ds}) yields the equation 
\begin{equation}\label{eq:back} \psi'' = - {(3\kp +F)F \over F_{\psi'}} \ ,  \end{equation} 
where $F(\psi') \ne 0$. Both equations reduce to ($\ref{eq:back}$), as they should as we have demanded that the equations are degenerate when imposing $h'=0$, $h=\kp$. Clearly the scalar field is evolving at this point, indicating that the de Sitter state is not a mathematical fixed point of the system. The dynamics of $\psi$ depends on the functional form of $F$ -- this reflects the fact that there is a family of functions $A$, $B$ that can screen $\rho_{\Lambda}$. If we fix $F(\psi')$ then we fix the $A$, $B$ functions appearing in the action. 
Note that a constant part of $A$ can be absorbed into 
$\rho_\Lambda$, and a constant part of $B$ gives rise to a 
total derivative term in the action, so we ignore both. 

One can calculate $B$ in terms of $F(\psi')$, 
\begin{equation} 
B(\psi') = -{c_{1}  \over 9 \kp^{2}} \left[ {3 \kp \over 3 \kp +F} + \ln\left({F \over 3 \kp +F}\right)\right] \ . \end{equation}
 
\noindent However, one cannot compute $A(\psi')$ in closed form without first specifying $F(\psi')$. 

Let us review our approach. We have imposed an ansatz $h=\kappa$ that is independent of the field $\psi$, $\psi'$, and $\rho_{\Lambda}$. For this choice to be a solution to the field equations, we require some form of redundancy. This can be achieved either by demanding that the scalar field equation is trivial at $h=\kappa$, or that the scalar field and scale factor equations are equivalent at $h=\kappa$. We are studying classes of models for which the latter condition is enforced, and find that any pair of $A,B$ functions related via Equations~(\ref{eq:at1}, \ref{eq:bt1}) will generate a degenerate de Sitter point. The function $F(\psi')$ is arbitrary and unphysical, it is simply a mechanism by which we can write the functions $A,B$ in parametric form. The second requirement for self-tuning, which is that the scalar field equation possesses a non-trivial $h'$ dependence, is also satisfied by these models.

To exhibit the self tuning behaviour described in this section, we provide the simplest non-trivial example of an action that admits de Sitter solutions and redundant field equations. In the Appendices we study some more complex models that also exhibit self tuning.

\subsection{Example: $\apsip=0$} \label{sec:consta} 

The simplest example that can be presented is one in which $A={\rm const}$, which value can be absorbed into the vacuum energy $\rho_{\Lambda}$. This model will be characterised only by the tadpole term $c_{1}M^{4} \psi$ and 
$M^4 B(\psi') \Box \psi$ in the action (in conjunction with the standard Einstein-Hilbert action). From Equations~(\ref{eq:at1}, \ref{eq:bt1}) a function $F(\psi')$ exists for which $A_{\psi'} = 0$: 
\begin{eqnarray} & & {c_{1} \over 3\kappa + F} + 3 \kappa c_1{F_{\psi'}\psi' \over F(3\kappa + F)^{2}} = 0 \\ 
& & F(\psi') = -{3\kappa \over 1-\psi'} \\
\label{eq:exa2} & & A(\psi') = {\rm const} \\ 
\label{eq:exb2} & & B(\psi') = {c_{1} \over 9 \kappa^{2}}\left( \ln \psi' + {1 \over \psi'} \right) \ . 
\end{eqnarray} 

The field equations read 
\begin{eqnarray}\label{eq:ex1_0} & & 3 {M_{\rm pl}^{2} \over M^{2}} h^{2} = {M_{\rm pl}^{2}M_{\Lambda}^{2} \over M^{4}} - c_{1} \psi + {c_{1} \over 3 \kappa^{2}}h (\psi' - 1) \\ 
\label{eq:ex1_1} & & 2 {M_{\rm pl}^{2} \over M^{2}}h' = {c_{1} \over 9\kappa^{2}}\left( 1 - {1 \over \psi'} \right) \left(\psi'' - 3h\psi' \right) \\ 
\label{eq:ex1_2} & & 0 = -c_{1} + {c_{1} \over 3 \kappa^{2}}\left(1 - {1 \over \psi'} \right) \left( h'   + 3 h^{2}  \right) + {c_{1}  \over 3\kappa^{2}} h  {\psi'' \over (\psi')^{2}} \ . 
\end{eqnarray} 
Equations~(\ref{eq:ex1_0}--\ref{eq:ex1_2}) describe the dynamics of the model specified by functions (\ref{eq:exa2}, \ref{eq:exb2}), for all $h$ and $\psi$. If we search for de Sitter solutions by fixing $h={\rm const}$, $h'=0$ we find that a solution exists for which $h = \kappa$ and both 
Equations~($\ref{eq:ex1_1},\ref{eq:ex1_2}$) reduce to 
\begin{equation}  \psi'' = 3 \kappa \psi' \ , 
\end{equation} 
that is, a de Sitter solution for which the field $\psi$ does not relax to a constant vacuum expectation value but continues evolving, with 
$\psi'\sim e^{3\kp\tau}$. 

We already knew that this de Sitter solution with time varying $\psi(\tau)$ would exist, as any $A,B$ functions that are related via Equations~($\ref{eq:at1},\ref{eq:bt1}$) will yield a particular solution with $h =\kappa$, $\psi' = \psi'(\tau)$. This solution is just one particular `critical point' of the system -- away from this de Sitter point the dynamics will depend on our choice of $A,B$.  The dynamics of $\psi(\tau)$ at the de Sitter point $h = \kp$ also depends on our choice of functions $A,B$. It is interesting that this very simple example works; a greater variety of models with different asymptotic behaviors can be found in Appendix~\ref{sec:apx}.

\subsection{General Case}

For any functional forms $A$, $B$ that satisfy the relations ($\ref{eq:at1},\ref{eq:bt1}$), we can write the scale factor and scalar field equations in terms of the function $F(\psi')$ as 

\begin{eqnarray} & & 2 {M_{\rm pl}^{2} \over M^{2}} h' = - {c_{1} F_{\psi'}\psi' \over F(3\kappa + F)^{2}}\left[\psi'' + {F(3\kappa + F) \over F_{\psi'}} \right] + {3 c_{1} F_{\psi'}(\psi')^{2} \over F(3\kappa + F)^{2}} (h - \kappa) \ , \\ 
\nonumber & & 0 = - {3 c_{1} \psi' \over F(3\kappa + F)^{2}}(h-\kappa) \psi'' \left[ F_{\psi'\psi'} - {(F_{\psi'})^{2} \over F} - 2 {(F_{\psi'})^{2} \over 3\kappa + F} \right] - {3 c_{1} F_{\psi'} \over F(3\kappa+F)^{2}} (h-\kappa)\psi'' - {3c_{1} F_{\psi'}\psi' \over F(3\kappa + F)^{2}}h'  \\
& & \qquad - 9{c_{1} F_{\psi'}\psi' \over F(3\kappa + F)^{2}} h(h-\kappa) + {3c_{1} \over 3\kappa + F}(h-\kappa) - {c_{1} F_{\psi'} \over (3\kappa + F)^{2}} \left[ \psi'' + {F(3\kappa + F) \over F_{\psi'}} \right] \ .  \end{eqnarray}

\noindent We see explicitly that -- by construction -- this class of models possesses a particular on shell solution to the field equations for which $h = \kappa$ and $\psi'' = - F(3\kappa + F)/F_{\psi'}$.

We can eliminate the term proportional to $\psi'' + F (3\kappa + F)/F_{\psi'}$ from these equations, leaving an equation for $h$ that vanishes on-shell,  

\begin{eqnarray}\nonumber  2 {M_{\rm pl}^{2} \over M^{2}} h' &=& {3 c_{1} F_{\psi'}(\psi')^{2} \over F(3\kappa + F)^{2}} (h - \kappa)  + {3 c_{1} (\psi')^{2} \over F^2(3\kappa + F)^{2}}(h-\kappa) \psi'' \left[ F_{\psi'\psi'} - {(F_{\psi'})^{2} \over F} - 2 {(F_{\psi'})^{2} \over 3\kappa + F} \right]  \\
\label{eq:pertf} &\quad& + {3 c_{1} F_{\psi'} \psi' \over F^{2}(3\kappa+F)^{2}} (h-\kappa)\psi'' + {3c_{1} F_{\psi'}(\psi')^{2} \over F^{2}(3\kappa + F)^{2}}h'   + 9{c_{1} F_{\psi'}(\psi')^{2} \over F^{2}(3\kappa + F)^{2}} h(h-\kappa) - {3c_{1} \psi' \over F(3\kappa + F)}(h-\kappa) \ . \end{eqnarray}

\subsection{Perturbations around de Sitter vacuum} 

Equation~($\ref{eq:pertf}$) can be used to test if the de Sitter point $h=\kappa$ for any particular model is a stable attractor. Perturbing equation~($\ref{eq:pertf}$) around the de Sitter point as $h = \kappa + \delta h$, we find 

\begin{equation} \label{eq:pegen} \left[ 2 {M_{\rm pl}^{2} \over M^{2}} - {3c_{1} F_{\psi'}(\psi')^{2} \over F^{2}(3\kappa + F)^{2}}\right] \delta h' = {3c_{1} \psi' \over F(3\kappa + F)} \left[ -2 + {3 F_{\psi'}\psi' \over 3\kappa + F}\left( 1 + {\kappa \over F} \right) - {\psi' F_{\psi'\psi'} \over F_{\psi'}} + {\psi' F_{\psi'} \over F} \right] \delta h + 9{c_{1} F_{\psi'}(\psi')^{2} \over F^{2}(3\kappa + F)^{2}}(\delta h)^{2}  \end{equation}

\noindent where we have used $\psi'' = -F(3\kappa + F)/F_{\psi'}$ as we are interested in small perturbations from the de Sitter state, $\delta h = h-\kappa \ll 1$. Equation ($\ref{eq:pegen}$) has not been truncated in $\delta h$, as the cofactors involving $F, F_{\psi'}$, $F_{\psi'\psi'}$ multiplying $\delta h$ and $(\delta h)^{2}$ are time dependent functions that may be small.  Regardless, we find that perturbations from the de Sitter state are described by Bernoulli's equation ($\ref{eq:pegen}$), which generically admits an exact solution. For infinitesimal perturbations $\delta h$ we can linearize 
Eq.~($\ref{eq:pegen}$), in which case the condition that $\delta h$ is a decaying mode corresponds to  

\begin{equation}   {3c_{1} \psi' \over F(3\kappa + F)} \left[ -2 + {3 F_{\psi'}\psi' \over 3\kappa + F}\left( 1 + {\kappa \over F} \right) - {\psi' F_{\psi'\psi'} \over F_{\psi'}} + {\psi' F_{\psi'} \over F} \right] \left[ 2 {M_{\rm pl}^{2} \over M^{2}} - {3c_{1} F_{\psi'}(\psi')^{2} \over F^{2}(3\kappa + F)^{2}}\right]^{-1} < 0 \ , \end{equation} 
evaluated {\it on-shell\/}. 

We apply these results to our simple example, to study the evolution of
perturbations away from the de Sitter vacuum. For $A, B$ given in Equations~($\ref{eq:exa2},\ref{eq:exb2}$) we can write the auxiliary function $F(\psi') = -3\kappa /(1-\psi')$. For this model, Equation~($\ref{eq:pegen}$) can be written as 

\begin{equation}\label{eq:bernoulli} \left[ 2\mu + {c_{1} \over 9\kappa^{3}}(1-\psi')^{2} \right] \delta h' = {2c_{1} \over 3 \kappa^{2}} \psi' (1-\psi') \delta h - {c_{1} \over 3 \kappa^{3}}(1-\psi')^{2} \delta h^{2}  \ ,\end{equation} 
where $\mu\equiv M_{\rm pl}^2/M^2$, and $\psi' = c_{} e^{3\kappa \tau}$ for constant $c$. We can generate approximate solutions to this equation for certain limiting cases. For $(\psi')^{2} \gg \mu \gg 1$, $\delta h$ decays exponentially to zero,  
\be 
\delta h=C\,e^{-6\kp\tau} \ . 
\ee 

\noindent Alternatively, for $\delta h \ll 1$, $(\psi')^{2} \ll \mu$ we find $\delta h$ undergoes a period of slow roll in which $\delta h' \ll \delta h$, 

\begin{equation} \delta h \simeq \delta h_{\rm 0} \left( 1 - {c_{1} \psi' (\psi'-2) \over 18 \kappa^{3} \mu}\right) \end{equation}

\noindent for constant $\delta h_{0} \ll 1$. As $\psi'$ increases with $\tau$, $\delta h$ will decrease with time, but the time dependence of $\delta h$ is suppressed by a factor of $\mu^{-1} \ll 1$. We revisit this in greater detail in Section~\ref{sec:numerical}.

\section{Ghost and Laplace Stability Conditions} \label{sec:ghost}

Metric and scalar field perturbations for generalised scalar tensor theories have been studied in \cite{DeFelice:2011hq,DeFelice:2010as,DeFelice:2010pv,DeFelice:2010nf} and specifically for the action ($\ref{eq:scal1}$) have been constructed in \cite{Deffayet:2010qz,Kobayashi:2011nu}. We adopt the unitary gauge $\delta \phi = 0$ and take

\begin{equation} ds^{2} = -(1+2\alpha)dt^{2} + 2 a^{2} \partial_{i}\beta dt dx^{i} + a^{2}(1+2{\cal R}_{\phi}) dx_{i}dx^{i} \ .\end{equation} 
We can eliminate $\alpha$ and $\beta$ using the constraint equations, 
leaving the following quadratic action for ${\cal R}_{\phi}$,  

\begin{equation} S^{(2)} = {1 \over 2} \int d\eta d^{3}x z^{2} \left[ {\cal G} ({\cal R}'_{\phi})^{2} - {\cal F} (\vec{\nabla}{\cal R}_{\phi})^{2} \right] \ , \end{equation} 

\noindent where $\eta$ is the conformal time, primes denote differentiation with respect to $\eta$, and $z, {\cal G}, {\cal F}$ are functions of the background quantities $\phi$, $H$ as 
\begin{eqnarray} 
& & z = {a\dot{\phi} \over H - G_{X} \dot{\phi}^{3}/(2M_{\rm pl}^{2})} \\ 
& & {\cal F} = K_{X} + 2 G_{X}\left( \ddot{\phi} + 2H \dot{\phi}\right) - 2 {G_{X}^{2} \over M_{\rm pl}^{2}}X^{2} + 2 G_{XX}X \ddot{\phi} - 2\left(G_{\phi} - X G_{\phi X} \right) \\ 
& & {\cal G} = K_{X} + 2X K_{XX} + 6 G_{X} H \dot{\phi} + 6 {G_{X}^{2} \over M_{\rm pl}^{2}}X^{2} - 2\left( G_{\phi} + X G_{\phi X}\right) + 6 G_{XX}HX\dot{\phi} \ . 
\end{eqnarray} 
The stability of the background spacetime requires ${\cal F} > 0$ (Laplace stability) and ${\cal G} > 0$ (ghost free condition). We can write 
these in terms of the $A$ and $B$ functions as 
\begin{eqnarray} 
{\cal F}\,\psi'&=& \apsip+\bpsip(\psi''+4h\psi')+\psi'\psi'' B_{\psi'\psi'} 
-\frac{M^2}{2M_{\rm pl}^2}\bpsip^2(\psi')^3 \label{eq:fscript}\\ 
{\cal G}&=& A_{\psi'\psi'}+3h\bpsip+3h\psi' B_{\psi'\psi'}+\frac{3M^2}{2M_{\rm pl}^2}\bpsip^2(\psi')^2 \ . \label{eq:gscript} 
\end{eqnarray} 

For the degenerate de Sitter background solutions characterised by Equations~($\ref{eq:at1},\ref{eq:bt1}$) and $\psi'' = -F(3\kappa+F)/F_{\psi'}$, we can write these functions as 

\begin{eqnarray} & & {\cal F}\psi' = {2 c_{1} \over 3\kappa + F} - {c_{1} \kappa F_{\psi'}\psi' \over F (3\kappa + F)^{2}} + c_{1} \psi' \left[ {F_{\psi'\psi'} \over F_{\psi'} (3\kappa+F)} - {F_{\psi'} \over F(3\kappa + F)} - {2 F_{\psi'} \over (3\kappa + F)^{2}}  \right] - {1 \over 2}{M^{2} \over M_{\rm pl}^{2}}{c_{1}^{2} (\psi')^{3} F_{\psi'}^{2} \over F^{2} (3\kappa + F)^{4}}     \\
& & {\cal G} = -{c_{1} F_{\psi'} \over (3\kappa + F)^{2}} + {3 \over 2}{M^{2} \over M_{\rm pl}^{2}} {c_{1}^{2} (\psi')^{2} F_{\psi'}^{2} \over F^{2} (3\kappa + F)^{4}} \ . \end{eqnarray}

Note that for $F_{\psi'}<0$ the ghost condition is always satisfied (with $c_{1} > 0$); otherwise 
the conditions will depend on the values of $c_1$ and $M/M_{\rm pl}$, as well as 
the form $F(\psi')$. 

For the model characterised by Equations~($\ref{eq:exa2},\ref{eq:exb2}$) these conditions can be written as 

\begin{eqnarray}\label{eq:calf} & & {\cal F}  = {c_{1} \over 9\kappa \psi'} \left( 4 - {1 \over \psi'}\right) - {c_{1}^{2} \over 162 \mu \kappa^{4}}\left( 1 - {1 \over \psi'}\right)^{2} \\
& & {\cal G} = {c_{1} \over 3 \kappa (\psi')^{2}} + {c_{1}^{2} \over 54 \mu \kappa^{4}}\left( 1 - {1 \over \psi'}\right)^{2} \ . \end{eqnarray}

\noindent The no-ghost condition ${\cal G}>0$ will be satisfied for all time, subject to $c_{1} > 0$. The Laplace condition ${\cal F} > 0$ will be  satisfied for a large region of $\psi'$ values, but ${\cal F}$ is not manifestly positive (and indeed is negative for $\psi'<1/4$). The second term on the right hand side of 
Equation~($\ref{eq:calf}$) is suppressed by a factor of $\mu^{-1}\equiv  M^{2}/M_{\rm pl}^{2} \ll 1$, but $\psi'$ grows monotonically and will make the first term asymptotically approach zero (for very large $\psi$ field values). Nevertheless, for large $\mu$ the region $1/4\le \psi'\le 72\kp^3\mu/c_1$ has a stable de Sitter vacuum. 

\begin{figure}
  \centering
  \includegraphics[width=0.8\textwidth]{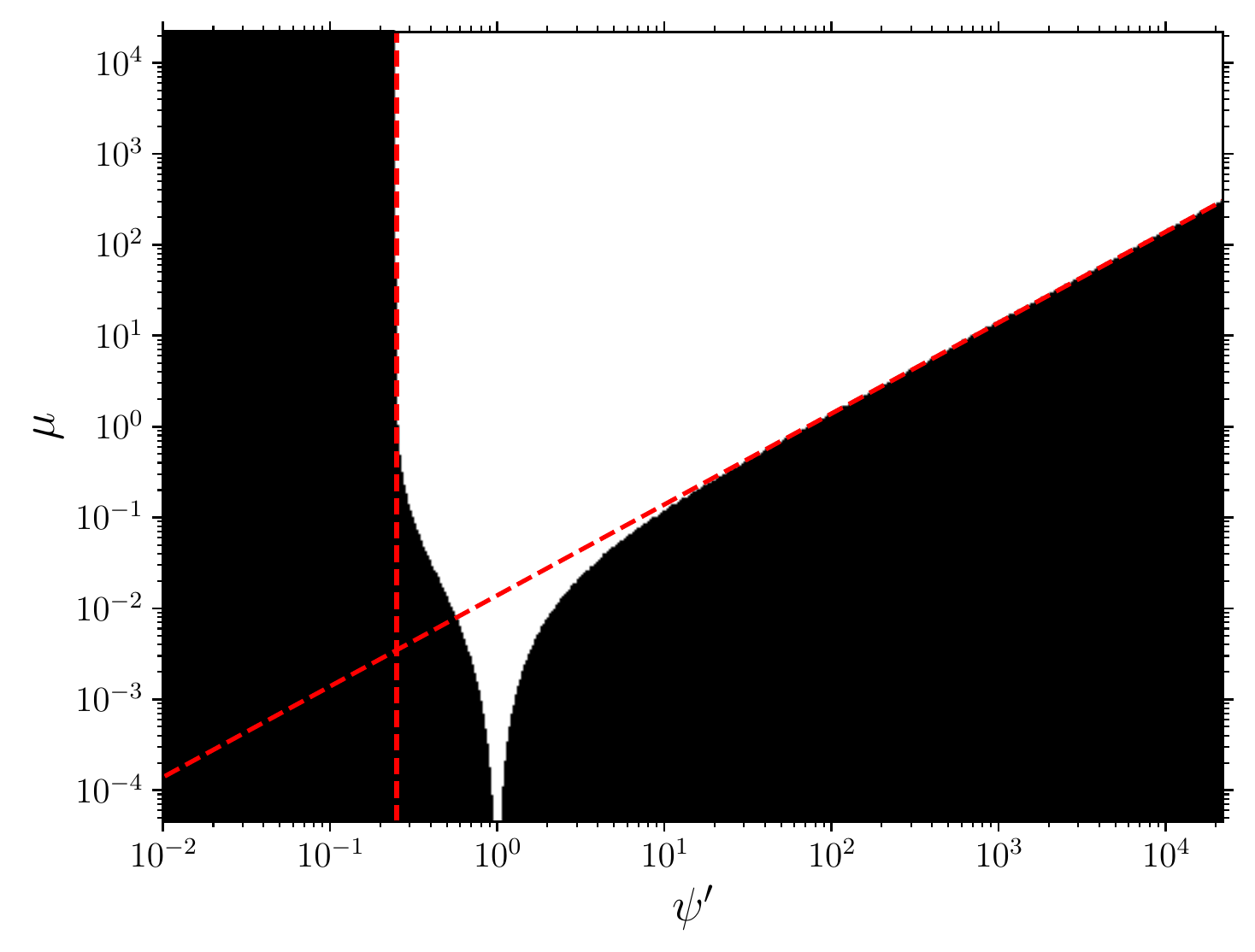} 
  \caption{Region of $(\psi',\mu)$ space for which the example model ($\ref{eq:exa2},\ref{eq:exb2}$) satisfies both the no-ghost ${\cal G} > 0$ and Laplace stability ${\cal F} > 0$ conditions (white). The black region corresponds to ${\cal F} < 0$, as ${\cal G}$ is manifestly positive for this model. The red dashed lines indicate the large $\mu$ asymptotic stability conditions $1/4 < \psi' < 72\kappa^{3}\mu/c_{1}$. 
 We have fixed $\kappa = 1$, $c_{1} = 1$.  }
  \label{fig:stability}
\end{figure}

In Figure~\ref{fig:stability} we exhibit the region of $(\psi',\mu)$ space for which both stability conditions ${\cal F} > 0$ and ${\cal G} > 0$ are satisfied (white region). The black region corresponds to Laplace unstable ${\cal F} < 0$. The red dashed lines correspond to the large $\mu$ stability limits $1/4\le \psi'\le 72\kp^3\mu/c_1$. For fixed $\mu$, there is a range of $\psi'$ values for which the de Sitter state is stable. In this numerical example we have fixed $\kappa = 1$, $c_{1} = 1$. Generally when $\psi'$ exceeds of order $\mu$ the stability is an issue, but note that for our actual universe with cosmic acceleration occurring at 
$M\sim H_0$ we expect $\mu\sim{\mathcal O}(10^{120})$ so this would be at a rather extreme value of $\psi'$. 

We note that violation of the so-called `Laplace instability' condition ${\cal F} > 0$ does not automatically condemn a model as non-viable, but rather indicates that the background spacetime is not stable, with exponentially growing perturbations. However, if the timescale associated with this instability is sufficiently large, the background plus perturbation split of the equations can still be performed over a suitable dynamical range.

\section{Numerical Solutions} \label{sec:numerical} 

To examine the stability properties of the de Sitter point in the non-linear regime - that is when $h, \psi'$ are a large distance in field space away from the critical point - we numerically evolve the dynamical equations of the example model ($\ref{eq:exa2},\ref{eq:exb2}$) derived in Section~\ref{sec:method}. (We consider further examples in Appendix~\ref{sec:apx}.) 
We select initial conditions for $h_{\rm i} = h(\tau_{\rm ini})$ and $\psi'_{\rm i} \equiv \psi'(\tau_{\rm ini})$, taking initial time $\tau_{\rm ini} = 0$. Following this, we fix $\psi_{\rm i} = \psi(\tau_{\rm ini})$ by solving the Hamiltonian constraint ($\ref{eq:h2}$). Then we evolve the dynamical equations ($\ref{eq:hdot},\ref{eq:psi2}$)
 for $h$, $\psi'$ over the time range $0 < \tau < \tau_{\rm f}$, with the end point $\tau_{\rm f}$ chosen arbitrarily. At each timestep we check that the Hamiltonian constraint remains solved to within the numerical tolerance that we select. This serves as a consistency check of our analysis. 

We fix the dimensionless free parameters in the problem as $\mu \equiv M_{\rm pl}^{2}/M^{2} = 10^{6}$, $M_{\rm pl}^{2} M_{\Lambda}^{2}/M^{4} = 10^{8}$, $c_{1}=1$, $\kappa = 1$. This choice presents a representative example in which there is a modest hierarchy between mass scales $M_{\rm pl} > M_{\Lambda} > M$. In reality we expect this hierarchy to be considerably larger if $M$ is the energy scale associated with the observed late time acceleration of the universe. 

\begin{figure}
  \centering
  \includegraphics[width=0.45\textwidth]{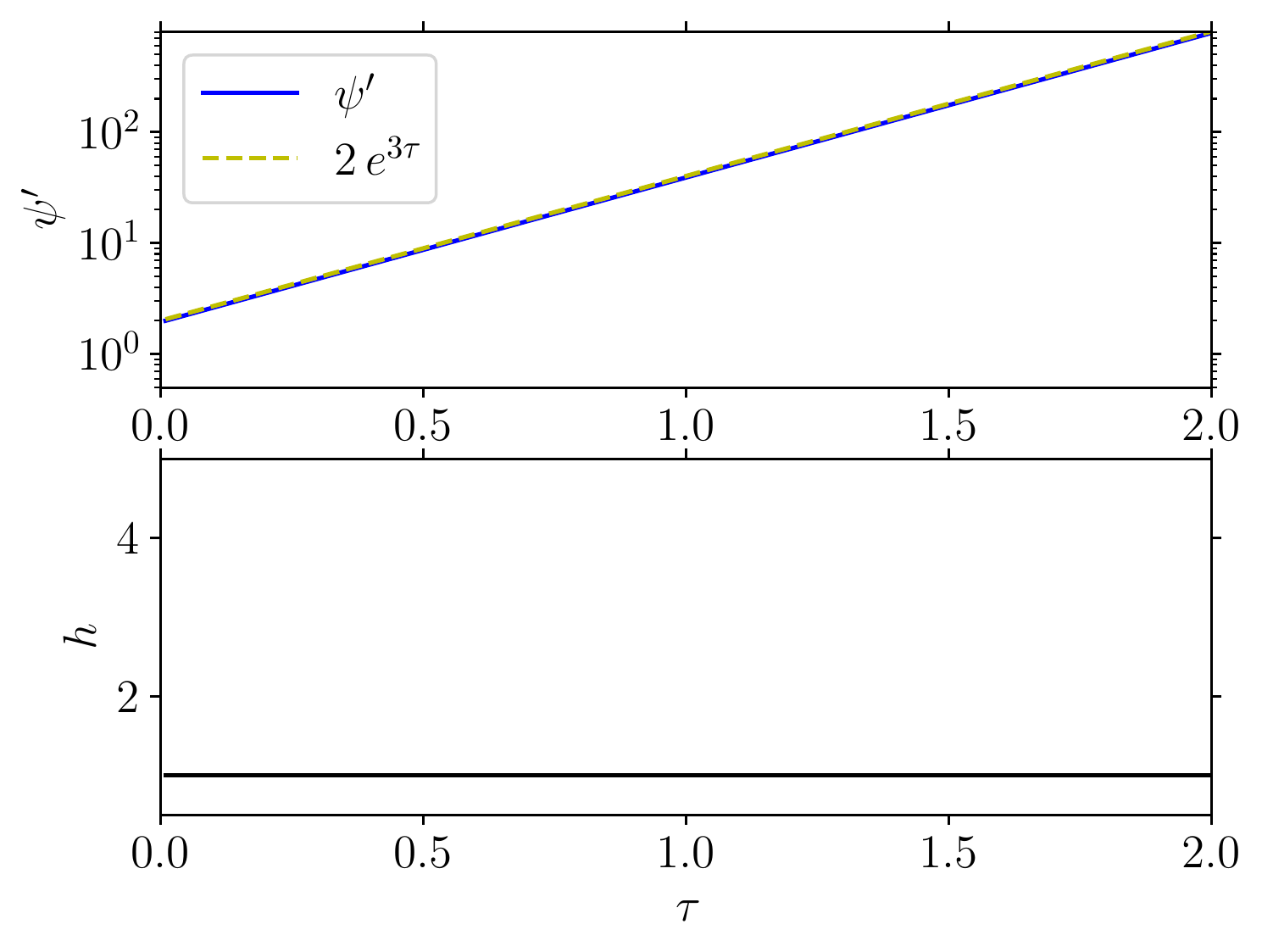} 
  \includegraphics[width=0.45\textwidth]{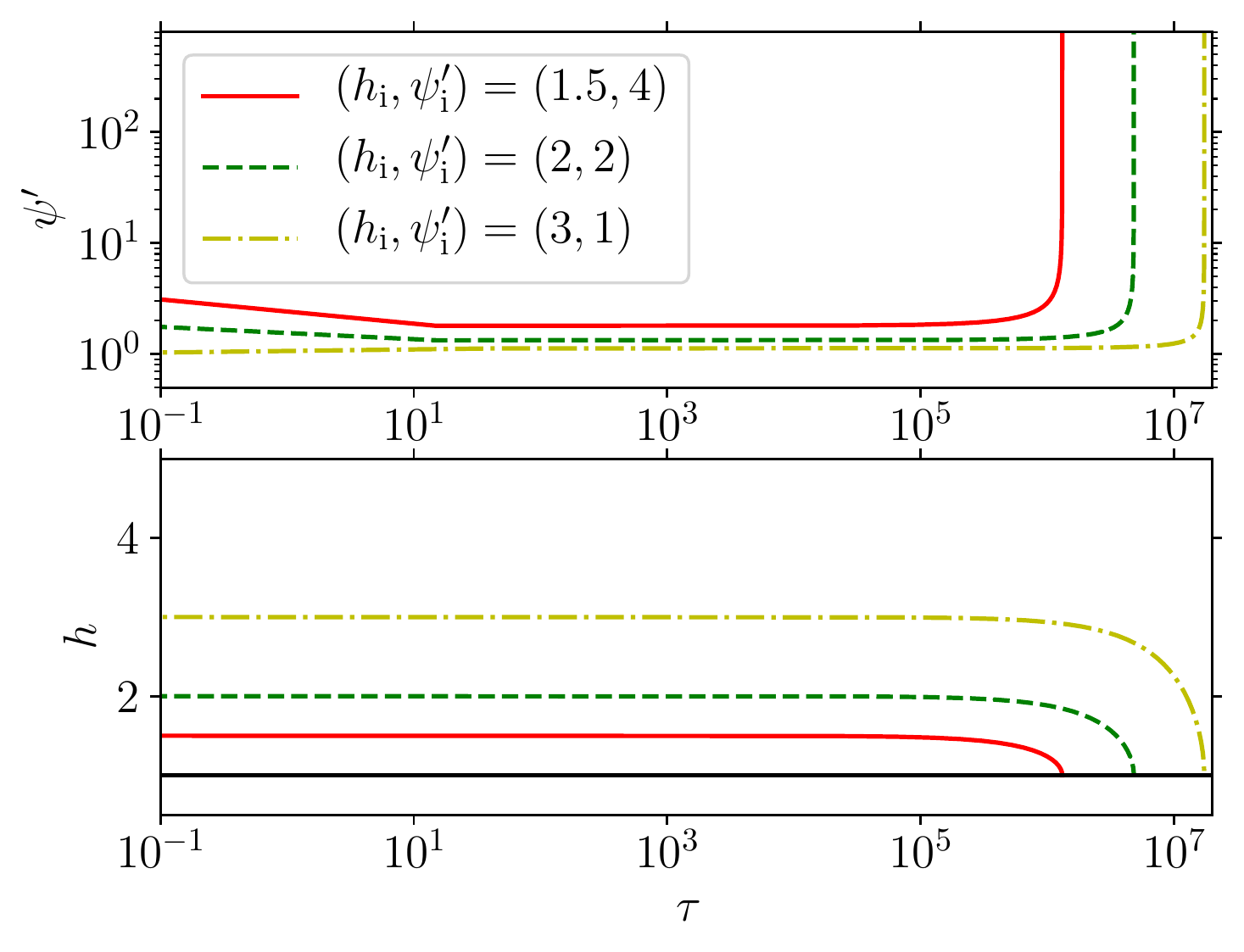}  \\ 
  \includegraphics[width=0.45\textwidth]{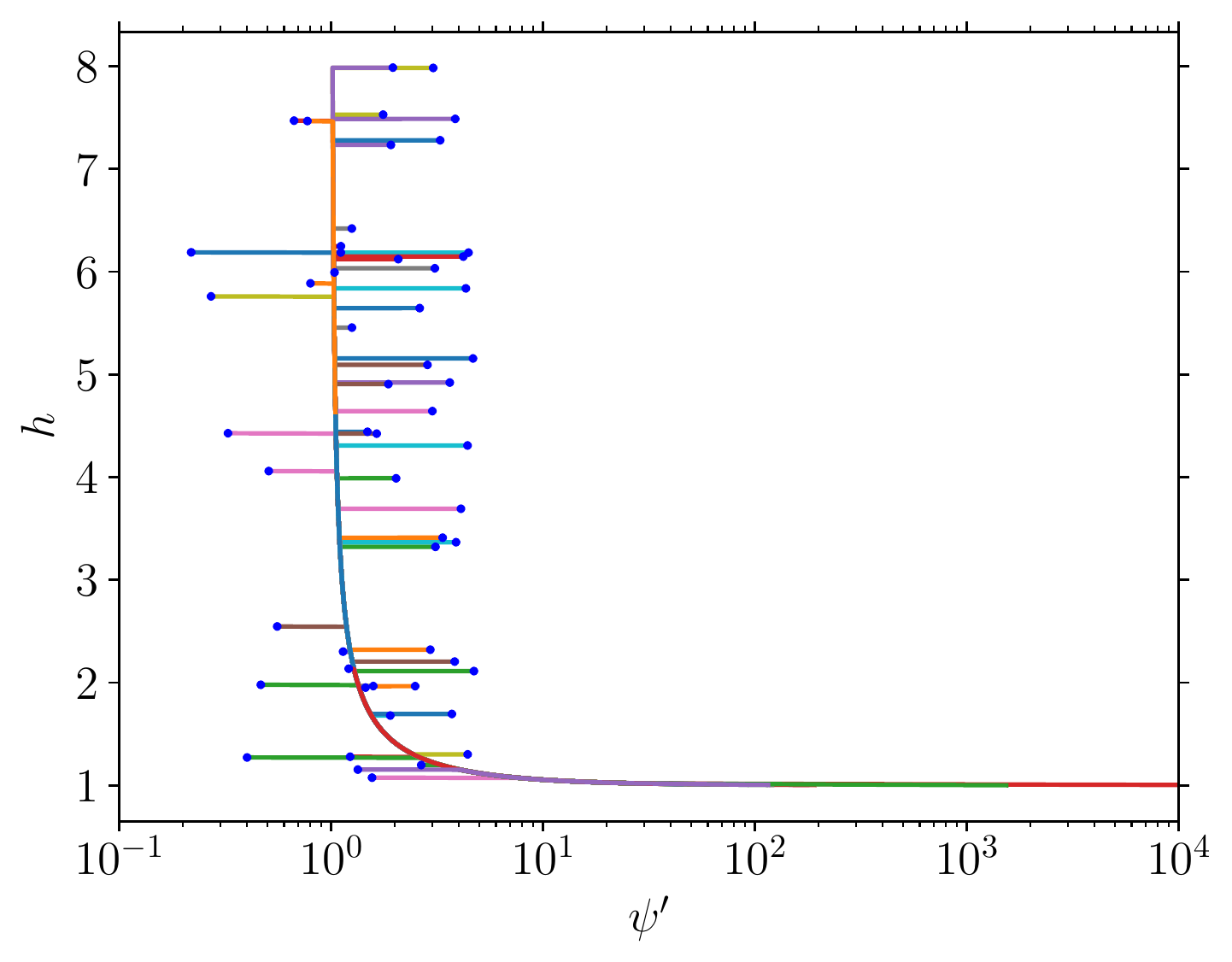}  
  \caption{ [Top left panel] The self-tuning solution is preserved, if $h$ starts at the self tuned value. We solve Equations~($\ref{eq:exa2},\ref{eq:exb2}$), fixing initial conditions $h_{i} = 1$, $\psi'_{\rm i}=2$, and parameters $\mu = 10^{6}$, $M_{\Lambda}^{2}/M^{2} = 10^{2}$, $\kappa = 1$, $c_{1}=1$. The Hubble parameter does not evolve -- this solution corresponds to an exact de Sitter state. The field $\psi$ stays dynamical, evolving indefinitely according to the analytic solution. [Top right panel] The self-tuning solution is reached, if $h$ starts away from the self tuned value. We move $h_{\rm i}$ away from the de Sitter point, fixing $(h_{\rm i},\psi'_{\rm i}) = (1.5, 4), (2,2), (3,1)$ for the red solid, green dashed and yellow dot-dashed lines respectively. 
The numerical solution is now characterised by two phases -- a slow roll, or loitering, period in which $h \simeq h_{\rm i}$ and $\psi' \simeq h_{\rm i}^{2}/(h_{\rm i}^{2} - \kappa^{2})$, and, after a time $\tau\sim\mu$, an approach to de Sitter with $h \to \kappa$ and $\psi' \to e^{3\kappa\tau}$.  [Bottom panel] The self-tuning solution is an attractor. Evolutionary tracks in the $(h,\psi')$ field space, starting the fields from random initial conditions in the range $1 < h < 8$ and $0 < \psi' < 5$, 
all converge to a single attractor trajectory. The blue dots correspond to the initial conditions, and the coloured tracks the subsequent evolution. One can observe periods of slow roll characterised by the horizontal tracks, followed by a common track in which $h \to 1$ and $\psi' \to \infty$. 
}
  \label{fig:3}
\end{figure}

In the top left panel of Figure~\ref{fig:3} we exhibit the numerical solutions for $\psi'$ and $h$ as a function of $\tau$, taking initial conditions $h_{\rm i} = 1$, $\psi'_{\rm i} = 2$. Since $h$ is starting at the on-shell solution, it does not evolve, but remains fixed at the de Sitter solution $h=\kappa$. The scalar field evolves at this point according to $\psi' = \psi'_{\rm i}\,e^{3\kappa\tau}$. This numerically confirms the existence of the self-tuning de Sitter point of the model. The de Sitter state is not a mathematical fixed point of the total dynamical system, as only $h$ relaxes to a constant expectation value while $\psi'$ stays dynamical. 

In the top right panel we plot the dynamics of $\psi'$, $h$ with initial conditions away from the de Sitter point. Specifically, the red solid, green dashed, and yellow dot-dashed curves represent the dynamics for initial conditions $(h_{\rm i}, \psi'_{\rm i})= (1.5, 4)$, $(2,2)$, $(3,1)$ respectively. Away from the de Sitter point, for this model the dimensionless Hubble parameter undergoes a period of slow roll from its initial value $h_{\rm i}$, with an eventual asymptote to $h \to \kappa$, while $\psi'$ grows exponentially on the approach $h \to \kappa$. The slow roll period arises due to the mass hierarchy that we have assumed, with $\mu = M_{\rm pl}^{2}/M^{2} \gg 1$. For this choice we can expand $h$ and $\psi'$ in powers of $\mu^{-1}$, arriving at  

\begin{eqnarray} & & \psi' \approx {h_{\rm i}^{2} \over h_{\rm i}^{2} - \kappa^{2}} + {\cal O}\left({1 \over \mu} \right) \\
& & h \approx h_{\rm i} - {c_{1} \over 6\mu} {h_{\rm i} \tau \over h_{\rm i}^{2} - \kappa^{2}} + {\cal O}\left({1 \over \mu^{2}}\right) \ . \end{eqnarray} 

The loitering period lasts until $\tau$ has become comparable to $\mu$, at which point $\psi'$ evolves quickly to its asymptotic exponential approach to 
de Sitter, and $h$ evolves quickly to its asymptotic constant value $h=\kappa$. 

In the bottom panel of Figure~\ref{fig:3} we select random initial conditions $0 < \psi'_{\rm i} < 5$ and $1 < h_{\rm i} < 8$  
and plot the dynamics of $(h, \psi')$ in field space for $N=55$ initial conditions. The blue points show the initial conditions $(h_{\rm i},\psi'_{\rm i})$ chosen randomly, and the coloured tracks represent the subsequent dynamics. The horizontal lines correspond to the slow roll period of the Hubble parameter. All trajectories approach a common dynamical track in the field space, on which $h \to \kappa$ and $\psi' \to \infty$. This indicates that the de Sitter point is an attractor.

\section{Existence of Matter Dominated Epoch} \label{sec:matter} 

Our goal is to not only attain a degenerate de Sitter state characterised by $h'=0$, $\psi = \psi(t)$, independent of the bare cosmological constant, but also to allow a viable expansion history away from the de Sitter state. This will involve suitably long periods of matter and radiation domination. One unfortunate aspect of the Fab Four \cite{Charmousis:2011bf,Charmousis:2011ea} and Fab Five \cite{Appleby:2011aa} actions for self tuning, 
for example, is that they screen not only the vacuum energy but also any matter and radiation present, making it difficult to obtain cosmologically interesting solutions\footnote{One can add scalar field potentials that restore a background 
expansion history that resembles matter or radiation \cite{Copeland:2012qf}.}.

\subsection{Tempered Tuning Preserves Matter} \label{sec:matdom} 

For the tempered self tuning considered in this work, a mechanism exists by which only the vacuum energy is screened. To observe this, we return to the original equations for our model, including a generic matter component, 

\begin{eqnarray} 
\label{eq:newham1} 3{M_{\rm pl}^{2} \over M^{2}} h^2&=& {\rho_{\rm mat} \over M^{4}} + {M_{\rm pl}^{2} M_{\Lambda}^{2} \over M^{4}} +\psi'\apsip-c_1\psi-A+3h\psip^2\bpsip \\ 
\label{eq:newham2} 2{M_{\rm pl}^{2} \over M^{2}}h' + 3{M_{\rm pl}^{2} \over M^{2}} h^2 &=& -{P_{\rm mat} \over M^{4}} + {M_{\rm pl}^{2} M_{\Lambda}^{2} \over M^{4}} -c_1\psi-A +\bpsip\psi'\psi'' \\ 
\label{eq:newham3} 0&=&3h\apsip+A_{\psi' \psi'}\psi''-c_1+3\bpsip\,(h'\psi'+h\psi''+3h^2\psi')+3hB_{\psi' \psi'}\psi'\psi'' \ .  
\end{eqnarray} 

When constructing degenerate models, our first step was to eliminate the $M^{2}_{\Lambda}$ term from Equation~($\ref{eq:newham2}$) by subtracting the Friedmann equation ($\ref{eq:newham1}$). We do this as we demand that Equations ($\ref{eq:newham2}$) and ($\ref{eq:newham3}$) are equivalent at $h=\kappa$ regardless of the value of $M_{\Lambda}$. By subtracting ($\ref{eq:newham1}$) from ($\ref{eq:newham2}$) we eliminate both the $M^{2}_{\Lambda}$ and $c_{1} \psi$ terms on the right hand side. When we introduce matter, performing the subtraction yields 

\begin{eqnarray}  
\label{eq:sub1} 2{M_{\rm pl}^{2} \over M^{2}}h' &=& - {\rho_{\rm mat} \over M^{4}} - {P_{\rm mat} \over M^{4}} -\apsip\psi' +\bpsip\psi'\,(\psi''-3h\psi') \\ 
\label{eq:sub2} 0&=&3h\apsip+A_{\psi' \psi'}\psi''-c_1+3\bpsip\,(h'\psi'+h\psi''+3h^2\psi')+3hB_{\psi' \psi'}\psi'\psi'' \ . 
\end{eqnarray}

\noindent We see that these equations can be equivalent only if $\rho_{\rm mat}+P_{\rm mat} = 0$. That is, this approach only screens the vacuum energy -- it automatically picks out vacuum energy for special treatment, unlike previous self tuning -- and both $h$ and $\psi$ will respond to $\rho_{\rm mat}$ and any energy density that does not have equation of state $w=P/\rho=-1$. If the matter is not directly coupled to $\psi$ (and it isn't in our model), then it should decay according to $\rho_{\rm mat} \propto a^{-3(1+w_{\rm mat})}$. This both gives a conventional matter dominated epoch and allows us to arrange the Equations~(\ref{eq:sub1}), (\ref{eq:sub2}) to be asymptotically equivalent as $\rho_{\rm mat} \to 0$. 

The above scenario is in contrast to the {\it on-shell\/} condition for models such as the Fab Four. In those models, the single, scalar field equation is selected to be trivially satisfied at the de Sitter point. This condition is independent of any other energy components, such as matter, that may be present. Regardless of a non-zero $\rho_{\rm mat}$ or any other component, at the de Sitter point one can always arrange the scalar field equation to be trivial. The redundancy condition proposed in this work -- that the two scalar field and scale factor equations are equivalent {\it on-shell\/} -- only applies to constant energy densities. The expansion rate $h$ will respond to any other energy contribution. 

Although $h$ will respond to a non-zero $\rho_{\rm mat}$ component, this does not automatically imply that the model will possess a viable expansion history for which $h^{2} \propto a^{-3}$ during pressureless matter domination. In this section we test the response of $h$, $\psi'$ to the presence of a non-zero matter component (specifically assuming pressureless matter). We re-write the field equations for our baseline model ($\ref{eq:exa2},\ref{eq:exb2}$) as 

\begin{eqnarray}\label{eq:wm0} & & 3 {M_{\rm pl}^{2} \over M^{2}} h^{2} = {\rho_{\rm mat} \over M^{4}} + {M_{\rm pl}^{2}M_{\Lambda}^{2} \over M^{4}} - c_{1} \psi + {c_{1} \over 3 \kappa^{2}}h (\psi' - 1) \\ 
\label{eq:wm1} & & 2 {M_{\rm pl}^{2} \over M^{2}}h' = - {\rho_{\rm mat} (1+w_{\rm mat}) \over M^{4}} + {c_{1} \over 9\kappa^{2}}\left( 1 - {1 \over \psi'} \right) \left(\psi'' - 3h\psi' \right) \\ 
\label{eq:wm2}  & & 0 = -c_{1} + {c_{1} \over 3 \kappa^{2}}\left(1 - {1 \over \psi'} \right) \left( h'   + 3 h^{2}  \right) + {c_{1}  \over 3\kappa^{2}} h  {\psi'' \over (\psi')^{2}} \\
\label{eq:wm3} & & \rho'_{\rm mat} + 3h \rho_{\rm mat} (1+w_{\rm mat})= 0 \ . 
\end{eqnarray}

We seek solutions in which the expansion rate closely mimics the standard matter era in general relativity, which would have the form 

\begin{eqnarray} & & 3 {M_{\rm pl}^{2} \over M^{2}} h_{\rm GR}^{2} = {\rho_{\rm mat} \over M^{4}} \\ 
& & 2 {M_{\rm pl}^{2} \over M^{2}} h'_{\rm GR} = -{\rho_{\rm mat} \over M^{4}} \\ 
& & \rho'_{\rm mat} + 3h_{\rm GR} \rho_{\rm mat} = 0 \ ,
\end{eqnarray} 
\noindent where we have taken $w_{\rm mat} = 0$. Therefore we take ansatz $h = h_{\rm GR} + \delta h$ where $h_{\rm GR} \gg 1$, $\delta h \ll h_{\rm GR}$. This limit corresponds to $\tau \ll 1$. To zeroth order in $\delta h$ we write the scalar field equation ($\ref{eq:wm2}$) as 

\begin{equation} {c_{1} \over 3 \kappa^{2}}h_{\rm GR} \psi'' + {c_{1} \over 3\kappa^{2}}\psi' (\psi'-1)(h'_{\rm GR} + 3 h_{\rm GR}^{2}) - c_{1} (\psi')^{2} = 0 \ , \end{equation}

\noindent which has solution 

\begin{equation}\label{eq:psipa} \psi'_{\rm a} = {2\tau \over \bar{c} + 2 \tau - 3 \kappa^{2}\tau^{3}} \ , \end{equation} 

\noindent for constant $\bar{c}$. In the small $\tau$ limit in which we are working, the density and pressure of the $\psi$ field can be approximated as 

\begin{eqnarray} & & \rho_{\psi} \simeq \rho_{0} - {2 c_{1} \over 9\kappa^{2} \tau} + {4 c_{1} \over 9\bar{c}\kappa^{2}} + {\cal O}(\tau) \\ 
& & P_{\psi} = -{c_{1} \over 9\kappa^{2} \tau} + {\cal O}(\tau) \ , \end{eqnarray} 

\noindent for constant $\rho_{0}$ (cf.~Eqs.~\ref{eq:ex1_0} and \ref{eq:ex1_1}). The energy density of the field is sub-dominant to $\rho_{\rm mat} \sim \tau^{-2}$, indicating that our initial ansatz $h \simeq h_{\rm GR} + \delta h$ to Equations~($\ref{eq:wm0}-\ref{eq:wm2}$) is valid (that is, such a solution exists). The scalar field energy density is initially sub-dominant but will grow relative to the matter component.

This model can exhibit a period of matter domination, in which the expansion rate is dominated by a dust component $\rho_{\rm mat}$. (And similarly a period of radiation domination.) 
As $\rho_{\rm mat} \to 0$, we must check numerically that the asymptotic behaviour of the dynamical system corresponds to a smooth transition to the de Sitter state $h \to \kappa$, and $\psi' \sim e^{3\kappa\tau}$. 

\begin{figure}
  \centering
  \includegraphics[width=0.98\textwidth]{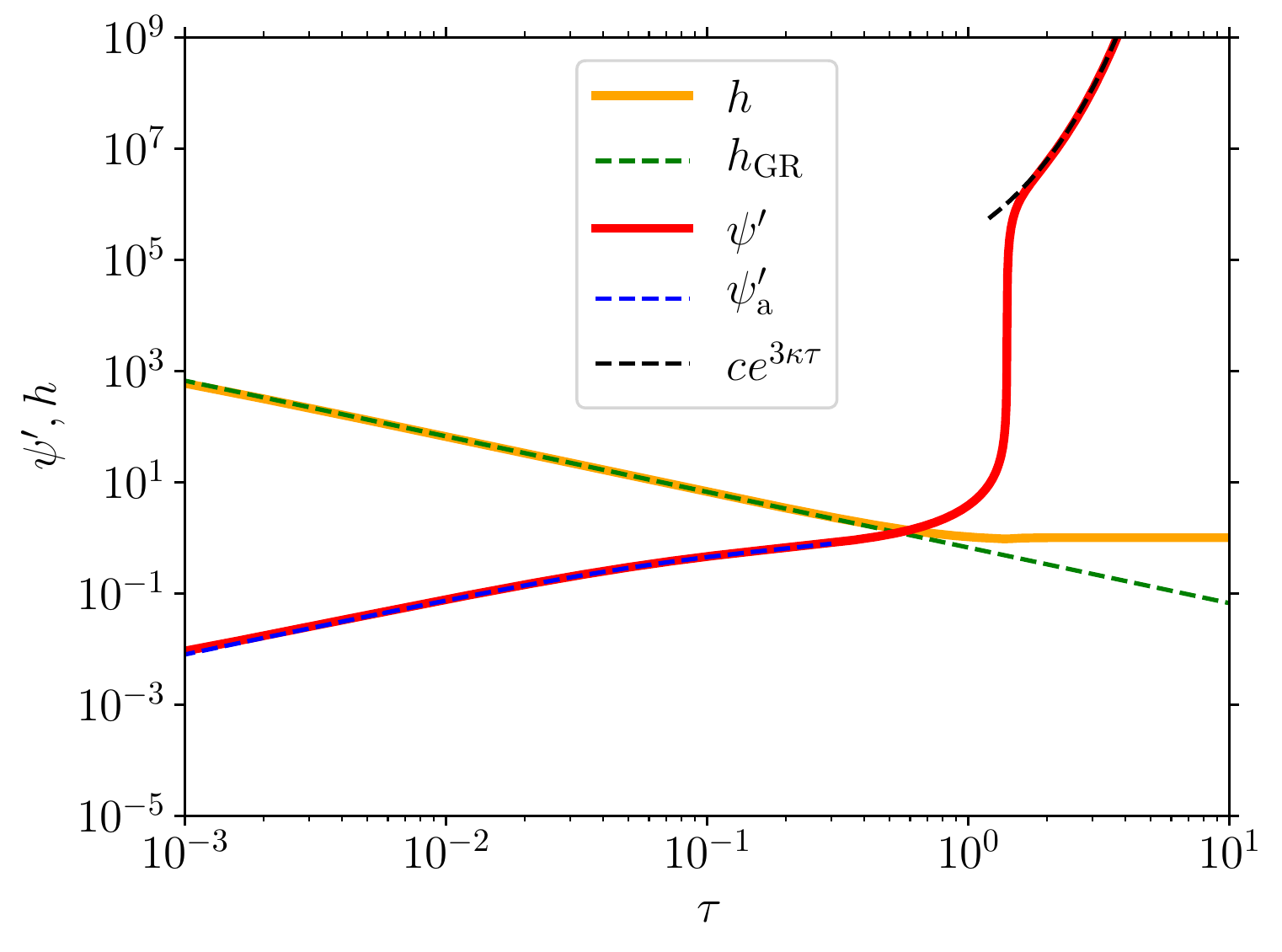} 
  \caption{The evolution of $h$ (solid yellow line) and $\psi'$ (red solid line), numerically evolved using Equations~($\ref{eq:wm1}-\ref{eq:wm3}$). $h$ exhibits a period of matter domination in which $h \simeq 2/(3\tau)$, followed by an approach to the degenerate de Sitter state $h \to \kappa$. The green dashed line indicates the GR solution $h_{\rm GR}=2/(3\tau)$. The cyan dot-dashed line is the analytic approximation $\psi'_{\rm a}$ given in Equation~($\ref{eq:psipa}$), and the black dashed line is $e^{3\kappa\tau}$, which is expected at the de Sitter point $h=\kappa$. 
  }
  \label{fig:matt}
\end{figure}

In Figure~\ref{fig:matt} we show the evolution of $h$ as a function of $\tau$, with $\tau_{\rm i}=10^{-3}$, $\tau_{\rm f} = 10$. We have fixed $\psi'_{\rm i} = 10^{-2}$ and $h_{\rm i} = h_{\rm GR, i}$. We set $c_{1} = 1$, $\kappa = 1$, $\mu = 10^{6}$ and $M_{\Lambda}^{2}/M^{2} = 10^{2}$.

The solid yellow curve is $h$ reconstructed numerically from the full equations, and the green dashed line is the matter dominated approximation $h_{\rm GR} = 2/3\tau$. 
The full solution follows the matter dominated GR solution for $\tau \ll 1$, as expected, before transitioning to accelerated 
expansion. The red solid line corresponds to $\psi'$ obtained by numerically integrating the full equations. The cyan dot-dashed line is the approximate solution ($\ref{eq:psipa}$), which closely matches the full solution for $\tau \ll 1$. The field $\psi'$ grows linearly with $\tau$, until a period of rapid growth as $h$ approaches $\kappa$. It then joins the exponential asymptote $\psi' \simeq c_{} e^{3\kappa\tau}$ (black dashed line) and $h \to \kappa$ asymptotically. 

Our numerical results indicate that the simplest model considered in this work can exhibit a period of matter domination that is preserved despite the self tuning and then gives way to an asymptotic approach to the well tempered degenerate de Sitter solution. Despite the presence of a large bare vacuum energy $M_{\rm pl}^2 M_{\Lambda}^{2}/M^{4} = 10^{8}$ in our numerical analysis, the background expansion rate $h$ is insensitive to this energy component.

\subsection{Self Tuning through a Phase Transition} \label{sec:phasetrans} 

One of the virtues of the tempered self tuning mechanism proposed in this work is that it only applies to constant energy densities -- $h$ will respond to any energy component with $w \neq -1$ and so a matter era is preserved. Therefore it is interesting to investigate how the self tuning reacts to a phase transition in the vacuum energy, e.g.\ as new particle degrees of freedom contribute, and the vacuum energy takes on different vacuum expectation 
values. Since during such a phase transition the vacuum energy $\rho_{\Lambda}$ will not be constant but will possess a non-trivial time dependence during the transitional period (and so during this time $w_{\Lambda} \neq -1$), it follows that the self tuning pauses and $h$ will vary during such events. That is, $h$  will be `kicked' from the {\it on-shell\/} solution $h=\kappa$. However, we expect that as $\rho_{\Lambda}$ approaches its new constant value (i.e.\ again $w_{\Lambda} \to -1$) then we have the restoration of self tuning, $h \to \kappa$, subject to this solution being an attractor.

\begin{figure}
  \centering
  \includegraphics[width=0.98\textwidth]{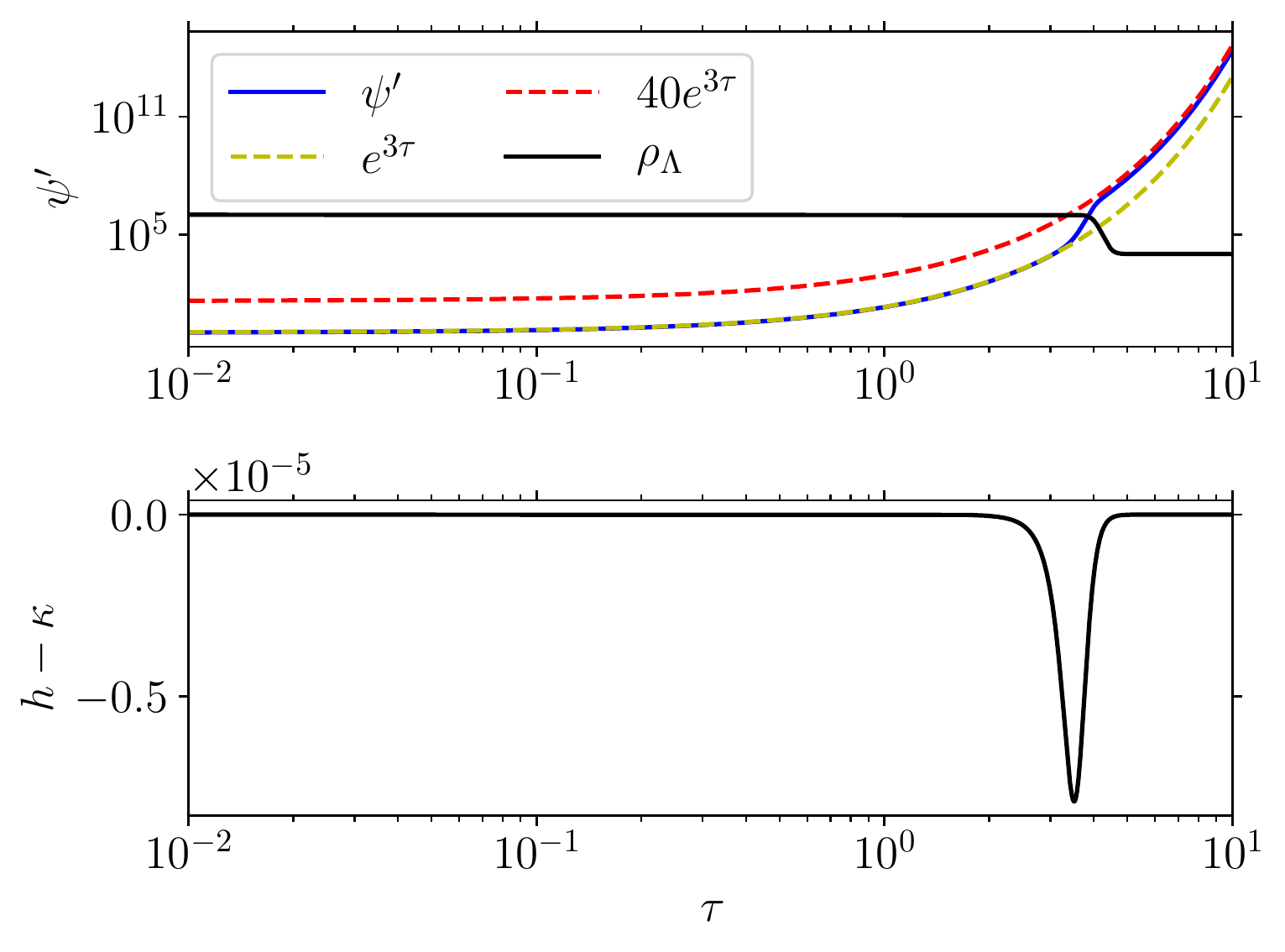}    
  \caption{If the vacuum energy undergoes a phase transition, the self tuning removes the cosmological constant on either side. We show the evolution of $h$ through the phase transition, from before, where $h$ begins {\it on-shell\/}, to after the phase transition. The blue solid line in the top panel is $\psi'$ and the black solid line is $\rho_{\Lambda}$, in which we have approximated a smooth phase transition as a tanh function with step at $\tau_{pt} = 4$. In the bottom panel we exhibit the deviation from the self tuned state, $h-\kappa$ (solid black line; note the $10^{-5}$ scale). Since during the phase transition the vacuum energy is not constant, then self tuning does not apply (recall this is what preserves a matter era) and $h$ evolves away from the de Sitter solution $h=\kappa$, but returns to this point (due to the dynamics of $\psi'$) when the vacuum energy $\rho_\Lambda$ is again constant. The shorter the phase transition, the briefer the period of deviation.} 
  \label{fig:pt}
\end{figure}

We demonstrate this behavior numerically in Figure~\ref{fig:pt} as the numerical solution to the following equations, 

\begin{eqnarray}\label{eq:pt0} & & 3 {M_{\rm pl}^{2} \over M^{2}} h^{2} = {\rho_{\Lambda} \over M^{4}}  - c_{1} \psi + {c_{1} \over 3 \kappa^{2}}h (\psi' - 1) \\ 
\label{eq:pt1} & & 2 {M_{\rm pl}^{2} \over M^{2}}h' = - {(\rho_{\Lambda} + P_{\Lambda}) \over M^{4}} + {c_{1} \over 9\kappa^{2}}\left( 1 - {1 \over \psi'} \right) \left(\psi'' - 3h\psi' \right) \\ 
\label{eq:pt2}  & & 0 = -c_{1} + {c_{1} \over 3 \kappa^{2}}\left(1 - {1 \over \psi'} \right) \left( h'   + 3 h^{2}  \right) + {c_{1}  \over 3\kappa^{2}} h  {\psi'' \over (\psi')^{2}} \\
\label{eq:pt3} & & \rho'_{\Lambda} + 3h (\rho_{\Lambda}+P_{\Lambda}) = 0 \ , 
\end{eqnarray} 

\noindent where we model a transition in $\rho_{\Lambda}$ numerically as 

\begin{equation} \rho_{\Lambda} = \rho_{1} - {(\rho_{1}-\rho_{2}) \over 2} \left[ 1 + \tanh \left({\tau - \tau_{\rm pt} \over \epsilon}\right) \right] \ , \end{equation} 

\noindent with $\rho_{1} = 10^{6}$, $\rho_{2} = 10^{4}$, $\tau_{\rm pt} = 4$ and $\epsilon = 0.2$. This evolves from an initial value $\rho_{\Lambda} \simeq \rho_{1}$ to $\rho_{\Lambda} \simeq \rho_{2}$ for $\tau > \tau_{\rm pt}$ in a smooth step. The effective pressure of this component is given by 

\begin{equation} P_{\Lambda} = -\rho_{\Lambda} - {\rho'_{\Lambda} \over 3h} \ .\end{equation}

We take initial conditions $\psi'_{\rm i} = 1$, $h_{\rm i} = \kappa$, and fix $\kappa = c_{1} = 1$. In the top panel $\psi'$ is shown as a solid blue line, and $\rho_{\Lambda}$ as a solid black line. At the point of transition in $\rho_{\Lambda}$, at $\tau = 4$, the field $\psi'$ changes magnitude (from yellow dashed to red dashed lines) but retains the self tuning time dependence $\psi' \sim e^{3\kappa\tau}$ asymptotically on either side of $\tau = \tau_{\rm pt}$. In the lower panel we exhibit the self tuning deviation $h-\kappa$, which begins at zero since $h_{\rm i} = \kappa$ but is kicked from this de Sitter state due to the transition -- because there is now a non-constant energy density that self tuning doesn't remove (as for matter). However, $h=\kappa$ is an attractor and the field returns to this point for $\tau > \tau_{\rm pt}$, i.e.\ self tuning reasserts itself after the phase transition and the bare cosmological constant remains canceled. 

Our numerical result indicates that phase transitions will affect the dynamics of the Hubble parameter, and a pressure singularity of the type considered in \cite{Charmousis:2011bf,Charmousis:2011ea} will not be screened for the period of the transition. The more rapid the transition, the briefer the effect on $h$. Thus, the tempered self tuning -- for the same reason that it preserves a matter era -- gives a (brief) signal of phase transitions. The  amplitude of the deviation in $h$ tends to be small, and it happens at early times, so detection would be challenging.

\section{Conclusions} \label{sec:concl} 

The original cosmological constant problem is one of the most puzzling aspects 
of gravitational and quantum physics, exacerbated by the fact that an extraordinarily 
small residual vacuum energy seems to be accelerating the 
universe currently. We have revisited the dynamical solution to this problem 
known as self tuning, taming several known issues and ending with a well tempered cosmological constant. 

Specifically, we construct a scalar-tensor model that possesses exact de Sitter solutions while the field remains dynamical, has the propagation speed of 
gravitational waves as the speed of light, removes a high energy cosmological constant while preserving a standard matter dominated expansion era, has the capacity 
for a nonlinear screening mechanism that gives general relativistic results for 
solar system tests, and is invariant under a shift symmetry $\psi \to \psi + c$ for controlling quantum corrections. The action is simply given by 
Eq.~(\ref{eq:scal1}) with functions (\ref{eq:kx}) and (\ref{eq:gx}). 

The desired de Sitter state with a self tuning, dynamical field requires some form of redundancy in the field equations, and we have identified two different redundancy conditions. One is the approach previously seen in the literature:  demanding that the scalar field equation vanishes identically at the de Sitter state (this is the ``trivial scalar'' approach discussed in Appendix~\ref{sec:appsfe}). We presented an alternative method that achieves 
the well-tempering, by imposing that the scale factor and scalar field equations are equivalent at the de Sitter point, and analyzed the general class of models that possesses this quality. 

A significant advantage of the well tempered approach is that the scalar field and scale factor equations can only be equivalent if the energy content has the 
characteristics of a vacuum energy, i.e.\ pressure equal to negative the energy 
density, or equation of state $w=-1$. Thus, this new class of models only self tunes vacuum energy, and matter content survives -- the Hubble parameter responds to the presence of dust and radiation and so one can have a conventional 
cosmic history of radiation domination, matter domination, then low energy cosmic acceleration despite the presence of a high energy cosmological constant. 
A future de Sitter state is approached asymptotically as an attractor. 

We have also shown that a large vacuum energy that undergoes phase transitions 
is successfully screened before and after, except for a brief period during the transition itself 
when $w\ne -1$. This may leave observational traces, though for the phase 
transition model we considered the duration and amplitude of the effect are small. It would be interesting to study further the limit as the phase 
transition becomes instantaneous, as used in \cite{Charmousis:2011bf,Charmousis:2011ea}. 

We presented a general prescription for generating evolutionary solutions for 
action functions $A(\psi')$ and $B(\psi')$, or the single auxiliary function $F(\psi')$. Numerical solutions for both a simple model with $A=\,$const and 
a variety of other behaviors agreed well with analytic results in the appropriate 
regimes (including an interesting loitering phase in the absence of matter). We 
verified stability of the de Sitter asymptote, demonstrating it was 
an attractor, and discussed the ghost free and Laplace stability conditions. In 
addition we translated our action functions to the property function approach of 
modified gravity from \cite{Bellini:2014fua}. 

Further avenues of future research could include more detailed studies of 
quantum corrections, as mentioned in \cite{Saltas:2016awg}, the nonlinear 
Vainshtein screening at work to evade fifth force constraints (cf.\ \cite{Babichev:2013cya,Charmousis:2014zaa,Charmousis:2015aya,Babichev:2015rva,Kaloper:2013vta,Appleby:2015ysa} for previous self-tuning models), and whether 
the form of the $\Box\phi$ term can be related to higher dimensional physics as 
in the original Dvali-Gabadadze-Porrati (DGP; \cite{Dvali:2000hr}) model. The last might 
offer some insight for solving the new cosmological constant problem, why the new mass 
scale $M$ in our action is such that acceleration occurs today. Note that our 
action is tightly determined by the shift symmetry condition, preventing the 
addition of a potential term. Conversely, one could view the tadpole term -- which 
plays a critical role -- as a 
linear potential, and $A(X)$ as a noncanonical kinetic term. 

The well-tempered self-tuning approach appears to be a rich mechanism leading to a 
valid solution to the old cosmological constant problem, shielding a high energy 
vacuum energy even through phase transitions and allowing a standard cosmic 
history. It offers new grounds for exploration of an exciting blend of 
quantum physics, gravity, and cosmology.

\acknowledgments 

This work is supported in part by the Energetic Cosmos Laboratory and by 
the U.S.\ Department of Energy, Office of Science, Office of High Energy 
Physics, under Award DE-SC-0007867 and contract no.\ DE-AC02-05CH11231.

\appendix  

\section{Further Examples} \label{sec:apx} 

The following models yield well tempered solutions with $h=\kp$. Depending on the functions $A,B$, the time dependence of $\psi$ will vary at the de Sitter point. In this appendix we construct functional forms of $A,B$ for which $\psi'$ exhibits power law or exponential time dependence at $h=\kappa$, either diverging or approaching $\psi' \to 0$ for $\tau \to \infty$.


First, let us look at $\psi'=c\tau^n$ at $h=\kappa$. The corresponding action functions $A,B$ that will give rise to this behaviour are 

\bea \label{eq:pla} 
\apsip&=&\frac{c_1}{3\kp}\,\left[1-b e^{-3\kp(\psi'/c)^{1/n}}\right]\, 
\left(1-\frac{3\kp}{n}\,(\psi'/c)^{1/n}\right) \\ 
\label{eq:plb} \bpsip&=&\frac{c_1}{3\kappa nc}\,\left[1-b e^{-3\kp(\psi'/c)^{1/n}}\right]\,
(\psi'/c)^{(1-n)/n} \ , 
\eea 

\noindent where $b$ is an arbitrary constant, indicating that there exists a one-parameter family of $A,B$ functions which can generate the same {\it on-shell} dynamics at $h=\kappa$. Note that for $n>0$, $\psi'$ diverges at 
late times, but all divergences are canceled in the equations of motion. Values $n<0$, i.e.\ inverse 
power laws, are also solutions; 
for $n<0$, $\psi'$ approaches zero for $\tau \to \infty$.

For an exponential dependence $\psi'=ce^{-3m\kappa\tau}$, the action functions are 
\bea 
\label{eq:plc}  \apsip&=&\frac{c_1}{3\kp}\frac{m+1}{m}\,\left[1-b\left(\frac{\psi'}{c}\right)^{1/m}\right] \\ 
\label{eq:pld}  \bpsip&=& -\frac{c_1}{9\kp^2m}\,\frac{1}{\psi'} \,\left[1-b\left(\frac{\psi'}{c}\right)^{1/m}\right] \ . 
\eea 
Note that $m=-1$ gives constant $A$, which corresponds to the example studied in 
Section~\ref{sec:consta}. Again, there exists a one-parameter ($b$)  family of functions that can generate particular $\psi' = c e^{-3m\kappa \tau}$ dynamics. The solutions are valid for either positive or negative $m$, with $\psi'$ 
approaching zero for positive $m$ and diverging for negative $m$. Again, 
all divergences cancel in the equations of motion. 

We evaluated numerically the behaviour of $\psi'$ and $h$ for four different models specified by the functions $A,B$, using $n=\pm1$ and $m=\pm1$, and fixing $b=1$, $\psi'_{\rm i} = 1$ at $\tau_{\rm i}=10^{-3}$, $\mu = 10^{6}$, $M_{\Lambda}^{2}/M^{2} = 10^{2}$. 
All these models indeed give a de Sitter state. If they start off the de Sitter state, the positive power law and exponential cases have behavior similar to that shown in Figure~\ref{fig:3}. The negative power law and exponential cases have ghosts. 

To ensure that the no-ghost condition is satisfied for these models we require $F_{\psi'}<0$, or $FB_{\psi'} > 0$. This condition is asymptotically satisfied as $\tau \to \infty$ for $m<0$ or $n > 0$. As mentioned in 
Section~\ref{sec:ghost}, the Laplace stability for these models will break down far in the future, when 
$\psi'\sim 10^{120}$, which may have interesting 
consequences.

\section{Trivial Scalar Field Equation {\it on-shell}} 
\label{sec:appsfe}

There are two methods by which one can construct de Sitter spacetimes that are insensitive to the value of the cosmological constant and for which the scalar field is not constant. Such a dynamical scenario requires some form of redundancy in the field equations -- this can be achieved either by forcing the scalar field and scale factor equations to be equivalent at the de Sitter state (the tempered approach, as discussed in the main body of the text) or by demanding that the scalar field equation is trivial at the de Sitter state \cite{Charmousis:2011bf,Charmousis:2011ea} (recall Section~\ref{sec:review}). For the latter case, the scalar field equation can be written as 

\begin{equation} \Upsilon(\lambda,h,\psi') \psi'' + \zeta(\lambda,h,h',\psi') = 0 \ ,\end{equation}

\noindent where $\Upsilon$, $\zeta$ are arbitrary functions that vanish identically for $h=\lambda$, $h'=0$ -- that is $\Upsilon(\lambda,\psi') = 0$, $\zeta(\lambda,\psi') = 0$. Two other conditions are required for self-tuning to occur: the Hamiltonian equation on-shell contains explicit $\psi'$ dependence and the scalar field equation contains explicit dependence on $h'$. 

Focusing on the action ($\ref{eq:scal1}$), the condition that there exists a de Sitter state $h = \lambda$, for constant $\lambda$, at which the scalar field equation is trivially satisfied requires the functions $A, B$ to obey the following relation, 
\begin{equation}\label{eq:sfev}  A_{\psi'} + 3 \lambda \psi' B_{\psi'} = {c_{1} \over 3\lambda} \ . \end{equation} 

\noindent This expression can also be satisfied without the tadpole term -- that is we can set $c_{1}=0$ and find solutions to Equation~(\ref{eq:sfev}). This is in contrast to the main body of the text, where it was found that $c_{1} \neq 0$ is required for the scalar field and scale factor equations to be equivalent at the de Sitter point (cf.~Equations~$\ref{eq:con1},\,\ref{eq:con2}$). 

In this appendix we propose a model that possesses multiple de Sitter states. Consider the following functions\footnote{These forms are  
motivated by solving Equation~($\ref{eq:back}$) implicitly as a function of time $\tau$, 
\begin{equation} 
F(\tau) = 3 \kp {c_{} e^{-3\kp\tau} \over 1 - c_{} e^{-3\kp\tau}} \ , \label{eq:ftau}
\end{equation} 
where $c$ is a constant, and then choosing $\psi'=c\,e^{-3\kp\tau}$, i.e.\ 
\be 
F(\psi')=3\kp\psi'/(1-\psi') \ . 
\ee 
However, Equation~(\ref{eq:ftau}) is only valid {\it on shell\/}, so the 
adoption of these particular forms for $A$ and $B$ for all times is purely heuristic, though guaranteeing the desired late time limit.}  

\begin{eqnarray}\label{eq:mod1} & & A(\psi') = {2c_{2} \over 3\kappa} \psi' \left( 1 - {\psi' \over 2}\right) \\
\label{eq:mod2} & & B(\psi') = {c_{2} \over 9\kappa^{2}} \left( \psi' - \ln \psi' \right) \ , \end{eqnarray} 

\noindent for arbitrary dimensionless constants $c_{2}$, $\kappa$. For this model, including the tadpole $c_{1}\psi$, the following dynamical equations can be derived, 

\begin{eqnarray}\label{eq:appex0} 
& & 3{M_{\rm pl}^{2}  \over M^{2}} h^{2} = {M_{\rm pl}^{2} M_{\Lambda}^{2} \over M^{4}} + {c_{2} \over 3 \kappa} (\psi')^{2}\left( {h \over \kappa} - 1 \right) - c_{1}\psi  - {c_{2} \over 3} {h \over \kappa^{2}} \psi' \\
\label{eq:appex1} & & 2 {M_{\rm pl}^{2}  \over M^{2}} h' = {2 c_{2} \over 3 \kappa} \psi' \left( \psi' - 1 \right) + {c_{2} \over 9\kappa^{2}} (\psi'-1) (\psi'' - 3h \psi') \\
\label{eq:appex2} & & {c_{2} \over 3\kappa} \left( {h \over \kappa} -2 \right)\psi'' + {c_{2} \over 3\kappa^{2}} (\psi'-1)h' + c_{2} \left( 2 {h \over \kappa} - {h^{2} \over \kappa^{2}}\right) - c_{1} - c_{2} \psi' \left( 2{h \over \kappa} - {h^{2} \over \kappa^{2}} \right) = 0 \ .\end{eqnarray} 

\noindent If we fix $c_{2} = c_{1}$, then this model possesses a de Sitter point $h=\kappa$ for which the scale factor and scalar field equations (\ref{eq:appex1}, \ref{eq:appex2}) are equivalent. Specifically, inserting the ansatz $h=\kappa$ into the above equations we find that both (\ref{eq:appex1}, \ref{eq:appex2}) reduce to 

\begin{equation} \psi'' = -3\kappa \psi' \ . 
\end{equation} 

\noindent This is an example of the tempered self-tuning discussed in the main body of the text. For $c_{2} = c_{1}$ this model satisfies the conditions ($\ref{eq:con1},\ref{eq:con2}$) for $h=\kappa$. For this solution the Laplace condition ${\cal F} > 0$ is asymptotically satisfied as $\tau \to \infty$ for $c_{2} > 0$ -- specifically ${\mathcal F}\psi' \to 2c_2/(9\kappa)$. 

However, this model also possesses a second de Sitter point that is only asymptotically reached, for which $h \to 2\kappa$ and $\psi'$ diverges linearly with $\tau$. Specifically, Equations~(\ref{eq:appex0}--\ref{eq:appex2}) possess the following solution for $\tau \gg 1$,  

\begin{equation} h = 2 \kappa + {1 \over 3\kappa \tau} + {\cal O}\left({1 \over \tau^{2}}\right), \qquad \qquad \psi' = {3\kp \over 2} \tau + {\cal O}(\tau^{0}) \ . 
\end{equation} 
This state does not possess the same redundancy property as the $h=\kp$ point -- the scalar field and scale factor equations are not equivalent if we insert the ansatz $h=2\kappa$. Regardless, it is an asymptotic de Sitter solution to the field equations.

However, if we set the tadpole term to zero $c_{1} = 0$ in Equations~(\ref{eq:appex0}--\ref{eq:appex2}), then $h=2\kappa$ is an exact de Sitter solution for which the scalar field Equation~(\ref{eq:appex2}) is identically zero. That is, the model (\ref{eq:mod1},\,\ref{eq:mod2}) possesses a ``trivial scalar'' self tuning solution $h=2\kappa$ in the absence of the tadpole term. For this solution, $\psi' = {\rm const}$ at $h=2\kappa$. When we re-introduce the tadpole term (fixing $c_{1} = c_{2}$) then the de Sitter state $h \to 2\kappa$ is present, but $\psi'$ now diverges linearly with $\tau$. 

More generally, for 
\bea 
A(\psi') &=& \frac{c_1}{3p\kp}\psi' +\frac{pc_2}{3\kp}\psi'\left(1-\frac{\psi'}{2}\right) \\ 
B(\psi') &=& {c_{2} \over 9\kappa^{2}} \left( \psi' - \ln \psi' \right) \ , 
\eea 
one reaches a de Sitter state with $h=p\kp$. This is a solution of 
the trivial scalar self tuning approach. Regarding the ghost condition, when $\apsip$ and 
$\bpsip$ are related through Equation~(\ref{eq:sfev}) to give the trivial 
scalar field equation property, then the de Sitter state is always ghost free. 

Let us clarify the relation between the tempered self tuning of the main 
text and the trivial scalar field equation self tuning. For a given 
action function $B$, one can compare the functions $A$ in the tempered 
and trivial scalar field equation approaches. This shows that the 
tempered approach is more general for $h=\kp$, reducing to the trivial 
scalar field equation result in the de Sitter state only when $F\to0$, 
$F_{\psi'}\to0$ in the approach to de Sitter. 
However, when $h=p\kp$ for $p>1$ then the trivial scalar field equation approach is the only means of 
attaining a ghost free state, and it can do so without the tadpole term. 

Specifically, consider whether a model exists that contains both types of redundancy -- that is a de Sitter state $h = \kappa$ for which the scalar field and scale factor equations are equivalent (tempered approach), and a de Sitter state $h=\lambda$ for which the scalar field equation vanishes identically (trivial scalar approach). For such a model, the function $B$ must satisfy the following conditions 

\begin{eqnarray}\label{eq:red} & & 3 (\lambda - \kappa)\psi' B_{\psi'} = {c_{1} \left[ - 3(\lambda-\kappa) + F \right] \over 3\lambda(3\kappa+F)} \\
\label{eq:red2} & & B_{\psi'} = - c_{1}\,  {F_{\psi'} \over F(3\kp +F)^{2}} \ . \end{eqnarray}

\noindent Equivalence for $A_{\psi'\psi'}$, which enters 
the equations, also requires $F_{\psi'}\to0$. 
If $\lambda\ne\kappa$, then Equation~(\ref{eq:red}) with $F\to0$ tells us that 
$\bpsip<0$, implying $F_{\psi'}>0$, violating the ghost condition for the tempered approach. Thus, as stated above, the tempered approach is more general for the $h=\kappa$ de Sitter state, while 
the trivial approach allows $h>\kappa$ limits. 

One consequence of Equations~(\ref{eq:red},\,\ref{eq:red2}) is that $\lambda \neq \kappa$, otherwise $F = 0$. The two de Sitter states characterised by $h=\lambda$ and $h=\kappa$ can be arbitrarily close ($\lambda \to \kappa$), but there cannot exist a single de Sitter state in which both redundancy properties are satisfied. 

We summarize that any functional forms of $A, B$ that satisfy either the relation (\ref{eq:sfev}) or the relations (\ref{eq:at1}, \ref{eq:bt1}) will possess a de Sitter point for which $h={\rm const}$, $\psi = \psi(\tau)$. This provides a complete set of self-tuning solutions of the action ($\ref{eq:scal1}$), subject to the condition that the model preserves the shift symmetry $\psi \to \psi + c$. 

\bibliography{biblio}{}

\end{document}